\documentclass[aps,twocolumn,groupedaddress,nofootinbib,english,nobalancelastpage,floatfix,superscriptaddress,preprintnumbers]{revtex4-1}

\usepackage[latin1]{inputenc}
\pdfoutput=1
\usepackage{bm}
\usepackage{multirow, amsmath, amssymb, amsfonts, array}
\usepackage{color}
\usepackage{graphicx,cancel}
\usepackage{calc}
\usepackage{dcolumn}
\usepackage{mathrsfs}
\usepackage{url}

\newcommand{\de}{\mbox{d}}
\newcommand{\non}[0]{\nonumber \\}
\newcommand{\rmd}{\rm d}
\newcommand{\GeV}{{\rm\ GeV}}
\newcommand{\MeV}{{\rm\ MeV}}
\newcommand{\TeV}{{\rm\ TeV}}
\newcommand{\sigmav}{\langle \sigma v \rangle}

\def\slashb#1{\setbox0=\hbox{$#1$}#1\hskip-\wd0\dimen0=5pt\advance
        \dimen0 by-\ht0\advance\dimen0 by\dp0\lower0.5\dimen0\hbox
          to\wd0{\hss\sl/\/\hss}}

\begin{document}

\title{Solar $\gamma$-rays as a Complementary Probe of Dark Matter}

\author{Chiara Arina}
\affiliation{Center for Cosmology, Particle Physics and Phenomenology (CP3), Universit\'e catholique de Louvain, B-1348 Louvain-la-Neuve, Belgium}
\author{Mihailo Backovi\'c}
\affiliation{Center for Cosmology, Particle Physics and Phenomenology (CP3), Universit\'e catholique de Louvain, B-1348 Louvain-la-Neuve, Belgium}
\author{Jan Heisig}
\affiliation{Institute for Theoretical Particle Physics and Cosmology, RWTH Aachen University, D-52056 Aachen, Germany}
\author{Michele Lucente}
\affiliation{Center for Cosmology, Particle Physics and Phenomenology (CP3), Universit\'e catholique de Louvain, B-1348 Louvain-la-Neuve, Belgium}
\preprint{CP3-17-08}
\preprint{TTK-17-07}

\begin{abstract}
We show that observations of solar $\gamma$-rays offer a novel probe of dark matter in scenarios where interactions with the visible sector proceed via a long-lived mediator. As a proof of principle, we demonstrate that there exists a class of models which yield solar $\gamma$-ray fluxes observable with the next generation of $\gamma$-ray telescopes, while being allowed by a variety of current experimental constraints. The parameter space allowed by big bang nucleosynthesis and beam dump experiments naturally leads to mediator lifetimes sufficient to produce observable solar $\gamma$-ray signals. The model allows for solar $\gamma$-ray fluxes up to orders of magnitude larger compared to dwarf spheroidal galaxies, without reaching equilibrium between dark matter annihilation and capture rate. Our results suggest that solar $\gamma$-ray observations are complementary, and in some cases superior,  to existing and future dark matter detection efforts.
\end{abstract}

\maketitle

\section{Introduction}\label{sec:intro} 

Despite enormous experimental effort, conclusive evidence of particle dark matter (DM), and its microscopic properties remains elusive. Searches using underground detectors, ground- and space-based telescopes and colliders have resulted in useful limits on particle DM properties, while measurements of the cosmic microwave background (CMB) by the Planck Collaboration have provided its relic density with unprecedented precision~\cite{Ade:2015xua}. Yet, existing results provide little information on the mass scale associated with particle DM and the possible interaction strengths between ordinary and dark matter. Complementarity of the existing methods to probe DM is a priceless asset, but will ultimately fail to cover the full spectrum of viable DM scenarios.  Some approaches to DM detection will also encounter technological limitations in the foreseeable future, \emph{e.g.}~the so-called ``neutrino floor'' in the case of direct detection experiments~\cite{Billard:2013qya}. In the quest for DM discovery, we should hence always strive towards novel methods for DM detection which can replace or complement the existing search efforts. 

It has long been known that large celestial bodies of high mass density, such as our Sun, could serve as ``reservoirs'' of DM \cite{Steigman:1978vs,Press:1985ug,Gould:1987ir}. The idea is based on the simple assumption that DM interacts with ordinary matter (quarks in particular) via interactions other than gravity, implying that  DM from the Galactic halo scatters off the matter inside the Sun. The resulting DM energy loss leads to gravitational capture of scattered DM which then accumulates in the center of the Sun, where it is allowed to annihilate into Standard Model (SM) states. 

There are several reasons the Sun could be an attractive target for $\gamma$-ray searches of DM:

\textbf{Low backgrounds:} The Sun is a poor source of $\gtrsim \text{GeV}$ scale $\gamma$-rays, providing a very low background environment for DM searches. Solar dynamics is characterized by $\mathcal{O}({\rm MeV})$  scale processes which do not result in significant output in $\gtrsim \text{GeV}$ scale radiation, with two significant exceptions. First, solar flares can be energetic enough to produce $\mathcal{O}(\text{GeV})$ $\gamma$-rays via neutral pion decays~\cite{1992ApJ}, but these processes are localized in time and can easily be vetoed. 
Second, two processes continuously produce  gamma-rays in the vicinity of the Sun: \emph{(i)} inverse Compton scattering of cosmic-rays off solar photons and \emph{(ii)} hadronic interaction of cosmic-rays with the solar atmosphere. Fermi-LAT observations provided evidence for such emissions up to photon energies of $\sim 100 \GeV$~\cite{2011ApJ...734..116A,Ng:2015gya}. On the other hand, there is no measurement of $\gamma$-rays from the Sun at higher energies: the authors in Ref.~\cite{Zhou:2016ljf} provide an estimate of such emission, which will be mostly localized to narrow angular regions at the solar edge and could in principle be vetoed if $\gamma$-ray telescopes can efficiently resolve the Sun.

\textbf{Astrophysical uncertainties:} The DM density at the Sun position is known with a 20\% accuracy ($\rho_\odot \sim 0.4  \, \GeV\,  {\rm cm}^{-3} $)~\cite{Read:2014qva,Salucci:2010qr}. Additionally, the solar capture rate does depend on the long term history of the Sun,  which finishes an orbit around the Galactic center in about $2 \times 10^8$ years.  Going beyond the  assumption  that  the  DM  Galactic  halo  is isotropic  and  smooth,  throughout  its  journey  the Sun will cross overdense or underdense regions with respect to an averaged density, which will ultimately influence the capture rate.  This effect has been estimated in Ref.~\cite{Arina:2013jya} to modify the capture rate and the expected neutrino, or in our case, photon fluxes by roughly 30\%. The uncertainties on the local and averaged DM densities are nonetheless  lower than those affecting the DM density profile close to the Galactic center by at least one order of magnitude.

\textbf{Proximity to Earth: } The Sun is close to the Earth compared to the Galactic center or the dwarf galaxies, leading to significantly lower suppressions of $\gamma$-ray fluxes due to distance from the detector. This implies that solar observations could be sensitive to lower DM annihilation rates, provided the resulting radiation can escape the solar surface. 

\textbf{$\gamma$-rays trace the source:} Similarly to neutrinos, the propagation of $\gamma$-rays is not affected by the solar magnetic field. Hence the direction of solar $\gamma$-rays points directly at their source.

In the past, solar capture of DM has mostly been discussed in the context of neutrino fluxes~\cite{Silk:1985ax,Ritz:1987mh,Kamionkowski:1991nj,Bergstrom:1996kp,Gondolo:2004sc,Blennow:2007tw,Sivertsson:2012qj}. However, several papers studied signals of DM captured by the Sun in other species of cosmic-rays and (dark) photons, see {\it e.g.}~\cite{Batell:2009zp,Schuster:2009fc,Schuster:2009au,Bell:2011sn,Feng:2016ijc,Adrian-Martinez:2016ujo,Ardid:2017lry}. The authors of Refs.~\cite{Batell:2009zp,Schuster:2009fc} argued that DM captured by the Sun could annihilate into a sizable flux of $\gamma$-rays and $e^+e^-$ pairs. Furthermore, the authors of Ref.~\cite{Feng:2016ijc} analyzed AMS-02 signals of DM annihilation via long-lived dark photons which decay to electron-positron pairs. Their results showed that AMS-02 could probe dark photon models with TeV scale DM with light mediators and small kinetic mixing. Scenarios similar to~\cite{Schuster:2009fc}, which can produce cosmic rays from DM annihilation in the Sun, have been constrained by the Fermi-LAT collaboration~\cite{Ajello:2011dq}. By revisiting the DM capture and annihilation from the center of the Sun, the authors of Ref.~\cite{Sivertsson:2009nx} have shown that DM annihilation just outside the surface of the Sun, in the so-called DM halo around the Sun, may be more easily detected. However, the expected continuum $\gamma$-ray flux would be negligibly small and below the sensitivity of future astrophysical probes.
Very recently the authors of Ref.~\cite{Leane:2017vag} investigated the emission of solar neutrinos and $\gamma$-rays caused by DM annihilating into long-lived mediators in the Sun. They studied a broad range of annihilation channels and derived constraints on the spin-dependent scattering cross section.
 
However, at the moment the question of whether observations of solar $\gamma$-rays could provide information complementary or competitive to other DM searches has not been studied in detail, leaving doubt about the real utility of solar $\gamma$-ray observations in DM physics.
Here we demonstrate, as a proof of principle, that there exists a class of viable DM models which can (in the foreseeable future)  be probed by  solar $\gamma$-ray observations to a degree competitive and/or complementary to other existing 
DM searches. Our results provide motivation for utilizing the present  (Fermi-LAT~\cite{Atwood:2013rka} and HAWC~\cite{Abeysekara:2013tza}) and future generation of $\gamma$-ray observatories (HERD~\cite{Zhang:2014qga,Huang:2015fca} and LHAASO~\cite{Zhen:2014zpa,He:2016del}) to measure high-energy solar $\gamma$-rays.

For the purpose of illustration we consider a simplified DM model in which a Dirac fermion DM ($X$) field interacts with SM quarks via a mixed scalar-pseudoscalar mediator ($Y$), with interaction strengths proportional to the quark Yukawa couplings. Requiring observable solar $\gamma$-ray fluxes generically implies large $Y$ lifetimes, as well as large mass hierarchies between $X$ and $Y$. We identify regions of the model parameter space which are consistent with DM relic density and are not ruled out by any existing experimental results, including direct detection, indirect detection and flavor constraints. In addition, as we consider long-lived mediators, we ensure that the model does not suffer from limits associated with big bang nucleosynthesis (BBN) and the CMB. We then argue that although such models can be probed by the ton scale direct detection experiments, such as XENON1T~\cite{Aprile:2012zx} and LZ~\cite{Akerib:2015cja}, observations of solar $\gamma$-rays (using the next generation of $\gamma$-ray telescopes) would present a complementary probe of DM dynamics, competitive with and in some mass ranges much more promising than $\gamma$-ray searches in the Galactic center and in dwarf spheroidal galaxies. 

The rest of the paper is organized as follows. In Sec.~\ref{sec:mod}, we motivate and define the simplified model used throughout the paper, while in Sec.~\ref{sec:bounds} we discuss all the bounds constraining the model parameters. In Sec.~\ref{sec:solar} we predict the $\gamma$-ray flux  from DM annihilation in the Sun and assess its detectability in comparison with other existing and future DM searches. We present our conclusions in Sec.~\ref{sec:concl}.

\section{Simplified Model: Mixed Pseudoscalar Mediator}\label{sec:mod}

In order to motivate our simplified model, we begin with the discussion of model ingredients which should (generically) be present in order to produce signals in solar $\gamma$-rays.
For DM to be captured by the Sun, the coupling of DM to SM quarks $g_{q}$ has to be nonzero. At some level in perturbation theory, nonzero $g_{q}$ will  naturally induce couplings to photons, making this a sufficient condition to produce $\gamma$-rays from DM annihilation. In addition, for a $\gamma$-ray signature of DM annihilation in the center of the Sun to be observable on Earth, the necessary condition is that DM annihilates into states (mediators) long-lived enough to decay outside the Sun's surface and decay at most at the Earth's surface. 
We will discuss the latter requirement at length in Sec.~\ref{sec:kinematics}, while here we concentrate on the former. 

Consider for instance the regime in which the two DM particles, each of them of mass $m_X$, annihilate into a pair of on-shell mediators, each of them with mass $m_Y$. The condition that the mediators escape the Sun before decaying can be written in terms of the mediator lifetime in the boosted frame as
\begin{equation}
	\frac{1}{\Gamma_Y}\frac{m_X}{m_Y} \gtrsim R_{\odot} \implies \left( \frac{\Gamma_Y}{\rm GeV}\right) \left(\frac{m_Y}{m_X}\right) \lesssim 2.84 \times 10^{-25}\, , \label{eq:rsun}
\end{equation}
where $R_\odot = 6.96\times 10^5 $ km is the radius of the Sun, and the factor $m_X/m_Y$ represents the boost of $Y$ in the rest frame of the Sun. Here we assumed  the dominant DM annihilation channel to be $X\bar{X} \rightarrow YY$, which is generically true for the regime $m_X \gg m_Y$. The immediate implication of Eq.~\eqref{eq:rsun} is that a model which can be probed with solar $\gamma$-rays should feature $\Gamma_Y \ll 1 \GeV$ and/or a large hierarchy between $m_X$ and $m_Y$.  Requiring a small mediator width suggests that the couplings of the mediator to the states it can decay to should be $\ll1$.\footnote{Except in special cases, such as when $m_Y$ is finely tuned to the threshold for the production of the decay products, resulting in strong kinematic suppressions of the $Y$ width. }

A simplified model with  a fermionic DM $X$, and a  mixed scalar-pseudoscalar mediator $Y$ that also couples to SM quarks represents an example of a model which satisfies the above-mentioned requirements.  The interactions are described by the Lagrangian:
\begin{eqnarray}
	\mathcal{L} & = & g_q y_q \, \bar{q} \left[ {\rm cos} \,\alpha +  i\, {\rm sin}\, \alpha\, \gamma_5 \right] q \,Y \nonumber \\
			  & + &  g_X \, \bar{X} \left[ {\rm cos} \,\alpha + i \, {\rm sin}\, \alpha\, \gamma_5 \right] X\, Y\,, \label{eq:lagr}
\end{eqnarray}
where $y_q\equiv \sqrt{2} m_q / v_h$ is the quark Yukawa coupling with $v_h=246 \GeV$ and $m_q$ the quark mass. 

In this paper we will only consider scenarios where $m_Y \ll m_X$, with $m_Y \sim {\cal O}(100) \MeV$, in order to naturally exhibit $\Gamma_Y \ll 1\GeV$. The mediator will decay to a pair of photons with a branching ratio of 100\%, as decays into gluons and light quarks will be suppressed by the fact that they would kinematically not be able to hadronize into a pair of pions, as long as $m_Y \lesssim 2 m_\pi $. Using Package-X~\cite{Patel:2015tea}, we have computed the decay of $Y$ into a pair of photons as
\begin{widetext}
\begin{equation}\label{eq:widthy}
\Gamma_Y = \frac{9}{8}\frac{g_q^2 \alpha_e^2 m_Y}{\pi^3} \left[  {\rm cos}^2\alpha \left| \sum_q Q^2_q \frac{m_q}{v_h} F_S\left(\frac{m_Y^2}{4 m_q^2}\right) \right|^2   
				+ {\rm sin} ^2 \alpha \left| \sum_q  Q^2_q \frac{m_q}{v_h} F_P\left(\frac{m_Y^2}{4 m_q^2}\right) \right|^2   \right]\,,
\end{equation}
\end{widetext}
where  $Q_q$ are the quark charges, $\alpha_e = 1/1 37$ is the electromagnetic fine structure constant and
\begin{eqnarray}
	F_S(x)& \equiv& \frac{1}{x^{3/2}}\left[-x + (x-1) {\rm arctanh}^2\left(\sqrt{\frac{x}{x-1}}\,\right)\right]\,, \nonumber \\
	F_P(x) & \equiv & \frac{1}{x^{1/2}}\left[{\rm arctanh}^2\left(\sqrt{\frac{x}{x-1}}\,\right)\right]\ \, .\nonumber
\end{eqnarray}
We have verified that the above formulas are consistent with the existing literature for pure scalar or pure pseudoscalar mediators~\cite{Haisch:2015ioa,Arina:2016cqj,Dolan:2014ska} and references therein. The prescription of replacing the $u,d$ and $s$ quark masses with the pion mass $m_\pi$ and the kaon mass $m_K$ respectively has been shown to approximate well the calculation for $Y$ width from chiral perturbation theory \cite{Leutwyler:1989tn,Dolan:2014ska}. 

Equation~\eqref{eq:widthy} allows us to estimate the simplified model parameter space which can, in principle, be probed by solar $\gamma$-ray measurements. Couplings of $g_q \sim {\cal O}(10^{-4}) $ easily result in decays of the mediator outside the Sun for $m_X \sim 1000 \GeV$, whereas heavier $m_X$ allow for couplings of $g_q \gtrsim 10^{-3}$ to be explored. The same parameter region is rather insensitive to the precise value of the mixing angle $\alpha$. Further details on the decay width and lifetime of the mediator necessary to produce solar $\gamma$-rays are given in Sec.~\ref{sec:kinematics}.

\subsection{Dark Matter Annihilation Channels and Relic Density Requirements} \label{sec:sigann}

In the early Universe, the freeze-out of DM in the $m_X \gg m_Y$ regime is typically governed by the $t$-channel process $X\bar{X} \rightarrow YY$.  For $m_X > m_t$ the $s$-channel $X\bar{X}\rightarrow t\bar{t}$ can also be significant, depending on the hierarchy between the couplings $g_X$ and $g_q$ (as discussed below).
For $m_X \gg m_Y$ and $m_X \gg m_t$, the thermal averaged cross section, expanded to $\langle v \rangle^2$ order in DM velocity, is given by \footnote{Here we omit the annihilation channels to quarks other than the $t$ due to the small $y_t/y_q$ ratio, as well as the fact that we will mostly discuss the regime of $m_X \gtrsim m_t$. In our numerical computation, however, we do include in $\sigmav$ the contribution of lighter quarks and gluons for $m_X < m_t$. Furthermore, we use the full expressions for finite $m_t, m_Y$. } 
\begin{eqnarray}
	 \frac{1}{2}\sigmav(X\bar{X}\rightarrow YY) & = & \frac{g_X^4 {\rm sin}^2 2 \alpha }{64 m_X^2 \pi} \nonumber \\
						&	+ &  \frac{g_X^4\left( 3 + 8 {\rm cos} 2\alpha + 7 {\rm cos} 4 \alpha \right) \langle v \rangle^2}{1536 m_X^2 \pi }\,, \nonumber \\
	 \frac{1}{2}\sigmav(X\bar{X}\rightarrow t \bar{t}) \, \, \, & =& \frac{3 g_q^2 g_X^2 y_t^2 }{64 m_X^2 \pi}\left[ {{\rm sin}^2 \alpha} + \frac{1}{4}{\rm cos}2 \alpha \langle v\rangle^2 \right]\,, \nonumber \\
	\label{eq:annihilation}
\end{eqnarray}
where we explicitly add a factor of $1/2$ to account for the fact that DM is a Dirac fermion. 
As indicated by Eqs.~\eqref{eq:annihilation}, in the limit of pure scalar or pure pseudoscalar couplings, the process $X\bar{X} \rightarrow YY$ is $p$-wave suppressed, while the mixed scalar-pseudoscalar coupling induces a leading $s$-wave annihilation cross section. For cos$\,\alpha \ll \,1$, the process $X\bar{X} \rightarrow YY$ is mostly $p$-wave at the characteristic freeze-out velocity of $\langle v \rangle \sim 0.2\, $, while the smaller $s$-wave component will be dominant for characteristic DM velocities in the Sun, galaxies and the CMB.

The $X\bar{X} \rightarrow t\bar{t}$ process is $s$-wave in the case of a pure pseudoscalar mediator, while it is $p$-wave suppressed in the case of a pure scalar or mixed mediator such that $\tan\alpha < \langle v \rangle/\sqrt{4+4\langle v \rangle^2}$. The last condition implies that at low DM velocities, for any configuration in which the mediator is more pseudoscalar than scalar the $s$-wave term will be dominating over the velocity suppressed one.  As the escape of the mediator from the Sun typically requires small $g_q$, it is evident that at freeze-out the $X\bar{X} \rightarrow t\bar{t}$ annihilation process will be subdominant. 

\begin{figure}[t]
\begin{center}
 \includegraphics[width=1.0 \columnwidth]{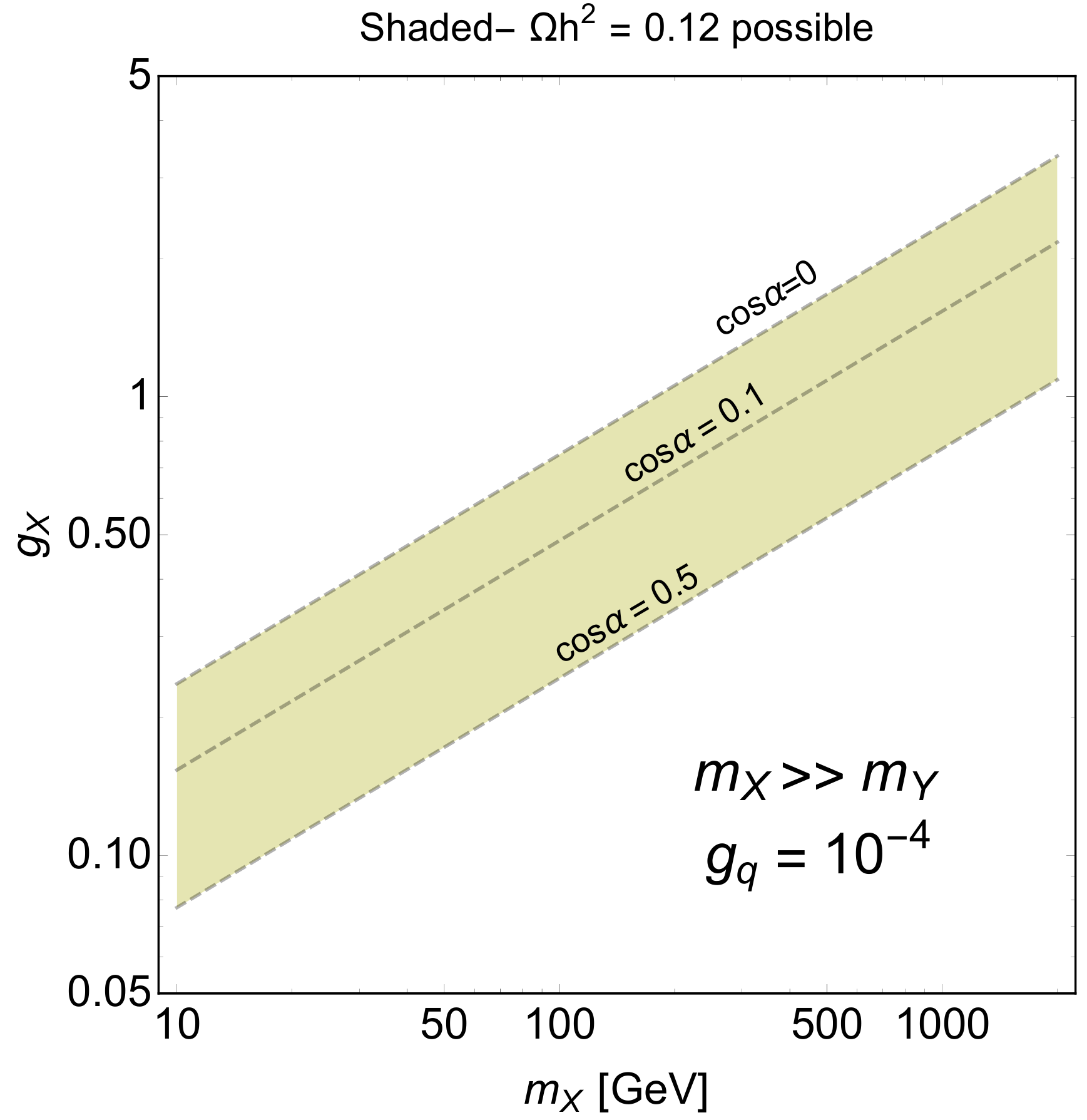}
\caption{Model parameter space that satisfies $\Omega h^2 = 0.12$. The dashed lines represent the lines of correct relic density for different values of $\cos\alpha$, as labeled.}
\label{fig:omegah2}
\end{center}
\end{figure}

Following the above considerations, $X\bar{X} \rightarrow YY$ is the dominant annihilation channel for fixing the DM relic density. The $X\bar{X} \rightarrow YY$ process is dependent on the size of the mediator coupling to DM and on the DM mass, but independent of the coupling of the mediator to quarks. This model feature partly decouples the requirements on obtaining the correct relic density from the calculation of other observables which involve quarks. Requiring $\Omega h^2=0.12$ implies $\langle\sigma v\rangle(X\bar{X}\rightarrow \rm YY)\approx3\times10^{-9} \GeV^{-2}$ at freeze-out, illustrated in Fig.~\ref{fig:omegah2}.

In the region $\cos\alpha \ll 1$, requiring correct relic density yields a simple condition:
\begin{equation}
	g_X \approx  \sqrt{\frac{m_X}{\GeV}} \left( 0.08 - 8.8\, {\rm cos}^2 \alpha + {\cal O}({\rm cos}^4 \alpha)\right)\,.\label{eq:omegah2}
\end{equation}
In the parameter space of interest (\emph{i.e.}~$m_X \sim 10 \GeV\! -\! 1 \TeV$), it is evident that $g_X \sim \mathcal{O}(0.1)\! -\! \mathcal{O}(1)$ is necessary, with the dependence on the mixing angle resulting in a maximum factor of $\sim 2$ difference on the required coupling.

We note that it is possible that the $X\bar{X} \rightarrow t\bar{t}$ process will become dominant over the $s$-wave $X\bar{X} \rightarrow YY$ term at lower DM velocity. Requiring the $X\bar{X} \rightarrow YY$ and $X\bar{X} \rightarrow t\bar{t}$  $s$-wave terms to be equal, implies quark-mediator couplings of the order of $g_q \sim 
 g_X \cos\alpha$. However, such a range of $g_q$ values either results in the mediator lifetimes insufficient to escape the solar surface or is constrained by several measurements that we discuss in the next section.  Hence, in the rest of the paper we consider the model parameter space in which the $X\bar{X} \rightarrow YY$ annihilation process is dominant at all DM velocities. 

\bigskip
\subsection{Elastic Scattering of Dark Matter off Nuclei} 
In the simplified model we consider there are four nonrelativistic operators~\cite{Fitzpatrick:2012ix,DelNobile:2013sia} that potentially contribute to the DM elastic scattering off a nucleon $n$: 
\begin{eqnarray}
\mathcal{O}_{\rm SI_1} & \propto &   \bar{X}X\,  \bar{n} n \,,\label{eq:siusual} \\ 
 \mathcal{O}_{\rm SI_2} & \propto &  \bar{X}\gamma_5 X\,  \bar{n} n  \,,\label{eq:si2}\\ 
\mathcal{O}_{\rm SD_1} & \propto &  \bar{X}X \, \bar{n} \gamma_5n  \, ,\label{eq:sd1}\\ 
 \mathcal{O}_{\rm SD_2} & \propto &  \bar{X}\gamma_5 X\,  \bar{n} \gamma_5 n \, . \label{eq:sd2} 
\end{eqnarray}
Equation~\eqref{eq:siusual} is the usual scalar spin-independent effective operator, while Eqs.~\eqref{eq:si2} and~\eqref{eq:sd1} come from the mixing between scalar and pseudoscalar couplings. Finally Eq.~\eqref{eq:sd2} is the well-known expression for pure pseudoscalar mediators. For fermionic DM, using the nonrelativistic spinor description, each $\gamma_5$ in the amplitude results in a momentum transfer squared ($q^2$) factor in the cross section, {\it e.g.}~$\bar{X}X\bar{n}\gamma^5 n$ is proportional to $\vec{q}\cdot\vec{s},$ where $\vec{q} \approx m_n \vec{v}$ and $\vec{s}$ is the nucleon spin vector. Hence the spin-independent part of the cross section is velocity independent and  proportional to $ \cos^2\alpha$. Conversely, parts of the DM-nucleon cross section which are proportional to the combination of the scalar and pseudoscalar coupling will be suppressed by $\langle v \rangle^2$ and $ \cos\alpha\sin \alpha$ with the coefficient of the same order as the scalar part. The pure pseudoscalar term of the DM-nucleon cross section will similarly be suppressed by $\langle v \rangle^4$ and $\sin^2\alpha$. For low enough DM and mediator masses, the momentum suppression in the numerator is compensated by the $m_Y^4$ factor in the denominator and the spin-dependent cross section can be comparable to the spin-independent cross section~\cite{Batell:2009zp,Arina:2014yna}. In our case, however, even a negligibly small admixture of a scalar component, $\cos\alpha \sim \langle v \rangle$, makes the usual scalar spin-independent operator of Eq.~\eqref{eq:siusual} dominant, as it is the only term which is not velocity suppressed.

Hence, unless $\cos \alpha \ll 10^{-3}$, the nucleon-DM scattering cross section is given by
\begin{widetext}
\begin{equation}
\sigma^{\rm SI}_{Xn} = \frac{\mu^2_n}{\pi}\frac{g_X^2 g_q^2 \cos^4\alpha\,  m_n^2}{m_Y^4} \left(\sum_{q = u,d,s} \frac{y_q}{m_q} f^n_q + \sum_{q=c,b,t} \frac{2}{27}\frac{y_q}{m_q} f^n_G\right)^2\,,  \label{eq:dd}
\end{equation}
\end{widetext}
where $\mu_n$ is the nucleon-DM reduced mass and $f^n_q,f^n_G$ are the quark and gluon content of the nucleons ($n$= neutron or proton) respectively.  For $u$ and $d$ quarks the uncertainties on the nucleon content values  are $\mathcal{O}(10\%)$, while for the $s$ quark the uncertainties rise up to a factor $\mathcal{O}(10)$. We fix $f^n_s = 0.043$ following~\cite{Junnarkar:2013ac}, which is a weighted average of lattice QCD calculations of $f^n_s$, and $f^n_u,f^n_d$ according to~\cite{Hoferichter:2015dsa}. Within uncertainties these values are consistent with 
results extracted from experimental information~\cite{Alarcon:2011zs,Alarcon:2012nr}. The gluon nucleon content is defined as $f^n_G = 1-\sum_{q=u,d,s} f^n_q$.

\subsection{Benchmarks Model Points}\label{sec:bench}

Here we are interested in a proof of principle that observations of solar $\gamma$-rays can provide competitive or complementary reach in DM searches. For this purpose we define several benchmark model points, summarized
in Tab.~\ref{tab:bench}, while we postpone a more detailed analysis of the full model parameter space for future work.
For each benchmark, we ensure that the parameter choice is consistent with the existing experimental constraints, which we discuss in detail in the following section.

\begin{table}[t]
{ \renewcommand{\arraystretch}{1.5}
\caption{Benchmark model points. We chose the values of $m_X$ to span a wide range of DM masses, while requiring the correct relic density fixes the value of $g_X$. All points give correct DM relic density and are compatible with the existing experimental constraints.\label{tab:bench} }
\begin{tabular}{cccccc}
Benchmark & $m_X [{\rm GeV}]$ & $m_Y [{\rm GeV}]$ & $g_X$ & $g_q$ & cos $\alpha$ \\
\hline\hline
 1a &  $10$ & $0.1$ & $0.24$ & $2 \times 10^{-5}$ & $0.01$ \\
 1b &  $10$ & $0.01$ & $0.24$ & $0.001$ & $0.001$ \\
 2a &  $100$ & $0.1$ & $0.76$ & $5 \times 10^{-5}$ & $0.012$ \\
 2b &  $100$ & $0.05$ & $0.76$ & $0.0001$ & $0.004$ \\
 3a &  $300$ & $0.1$ & $1.4$ & $0.0001$ & $0.01$ \\
 3b &  $300$ & $0.05$ & $1.4$ & $7 \times 10^{-5}$ & $0.004$ \\
 4a &  $1000$ & $0.1$ & $2.5$ & $9 \times 10^{-5}$ & $0.011$ \\
 4b &  $1000$ & $0.05$ & $2.5$ & $0.0002$ & $0.003$ \\
 5a &  $1800$ & $0.1$ & $3.4$ & $0.0001$ & $0.011$ \\
 5b &  $1800$ & $0.05$ & $3.4$ & $0.00012$ & $0.003$ \\
\hline
\end{tabular} }
\end{table}

\section{Existing Constraints and Future Sensitivity}\label{sec:bounds}

The mixed mediator scenario we study here provides signatures in a wide range of experiments, spanning cosmology, flavor physics, ground- and space-based DM searches. In the following we give a brief overview of all the relevant experimental constraints, and shortly discuss the ability of the future experiments to probe our simplified model. 

\subsection{Direct Detection}

Since we are considering the regime where scalar spin-independent scattering of DM off nuclei dominates, we here use only constraints on $\sigma_{Xn}^{\rm SI}$ from the LUX experiment~\cite{Akerib:2016vxi} at 90\% C.L. (confidence level). Note that we chose the scenario with dominant $\sigma_{Xn}^{\rm SI}$ purposefully, as an illustration of a model which is more difficult to survive all experimental constraints compared to a model which is dominated by spin-dependent scattering (subject to much weaker constraints from direct detection). 

\begin{figure}[b]
\begin{center}
 \includegraphics[width=1.0\columnwidth]{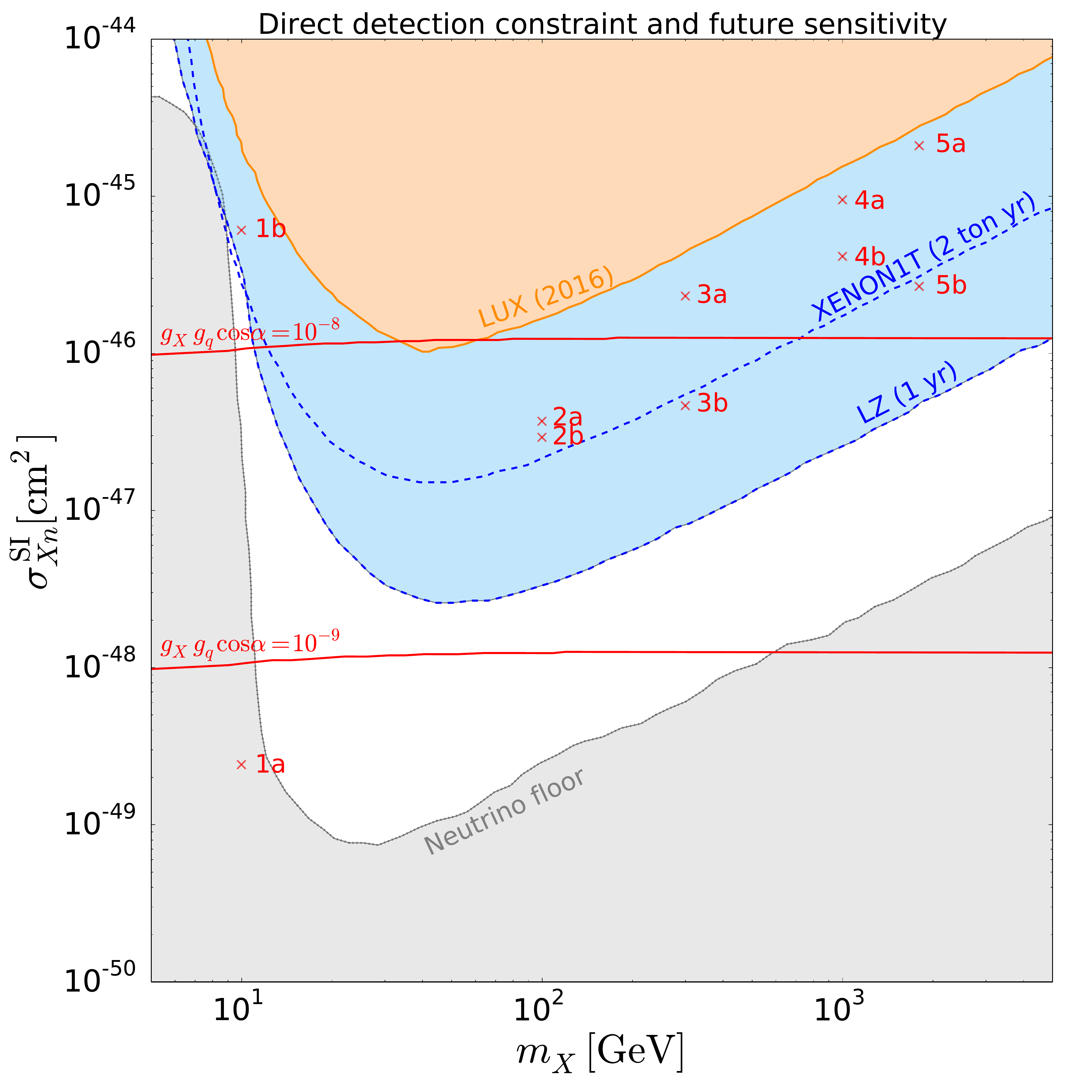}
\caption{Parameter space allowed by direct detection constraints. The thick solid curve represents the existing LUX bound on the spin-independent DM nucleon scattering  cross section at 90\% C.L.\@. The red solid lines show the respective cross sections in our model assuming various values for the product of the coupling parameters. The dashed curves show the projected bounds by the XENON1T~\cite{Aprile:2012zx} and LZ~\cite{Akerib:2015cja} experiments, 
while the gray line denotes the expected neutrino background~\cite{Billard:2013qya}. Excluded regions and regions in the reach of future experiments are shaded. We assume $m_Y = 100 \MeV$ for the purpose of the horizontal lines. The red crosses represent the benchmark points from Tab.~\ref{tab:bench}.}
\label{fig:dd}
\end{center}
\end{figure}

Figure~\ref{fig:dd} shows the portion of parameter space which is compatible with the existing LUX bound, as well as projections for the reach of XENON1T and LZ experiments. The red, horizontal lines show the magnitude of the spin-independent cross section in our model assuming a particular value for the product  $g_q g_X {\rm cos}^2 \alpha$, assuming $m_Y = 100 \MeV$ for illustration. Fixing $g_q \sim 10^{-4}$, roughly necessary for the mediator to decay dominantly outside the solar surface we find that $g_X{\rm  cos}^2 \alpha \lesssim 10^{-4}$ is allowed by the current LUX limit over a wide range of DM masses, and will be probed by XENON1T up to $m_X \sim 700 \GeV$ assuming a 2-year exposure. Conversely, the coupling product of $\lesssim 10^{-9}$ will not be efficiently probed by any direct detection experiment in the foreseeable future assuming  $m_Y\approx 100 \MeV$. Note that a part of the model parameter space gives rise to elastic DM-nucleon scattering cross sections smaller that the predicted neutrino-nucleus elastic scattering (denoted by the shaded gray region), giving rise to the irreducible neutrino background for DM searches (see {\it e.g.} the benchmark model point 1a).

\subsection {Observations of Dwarf Spheroidal Galaxies and the Galactic Center}

\begin{figure}[h]
\begin{center}
 \includegraphics[width=1.\columnwidth]{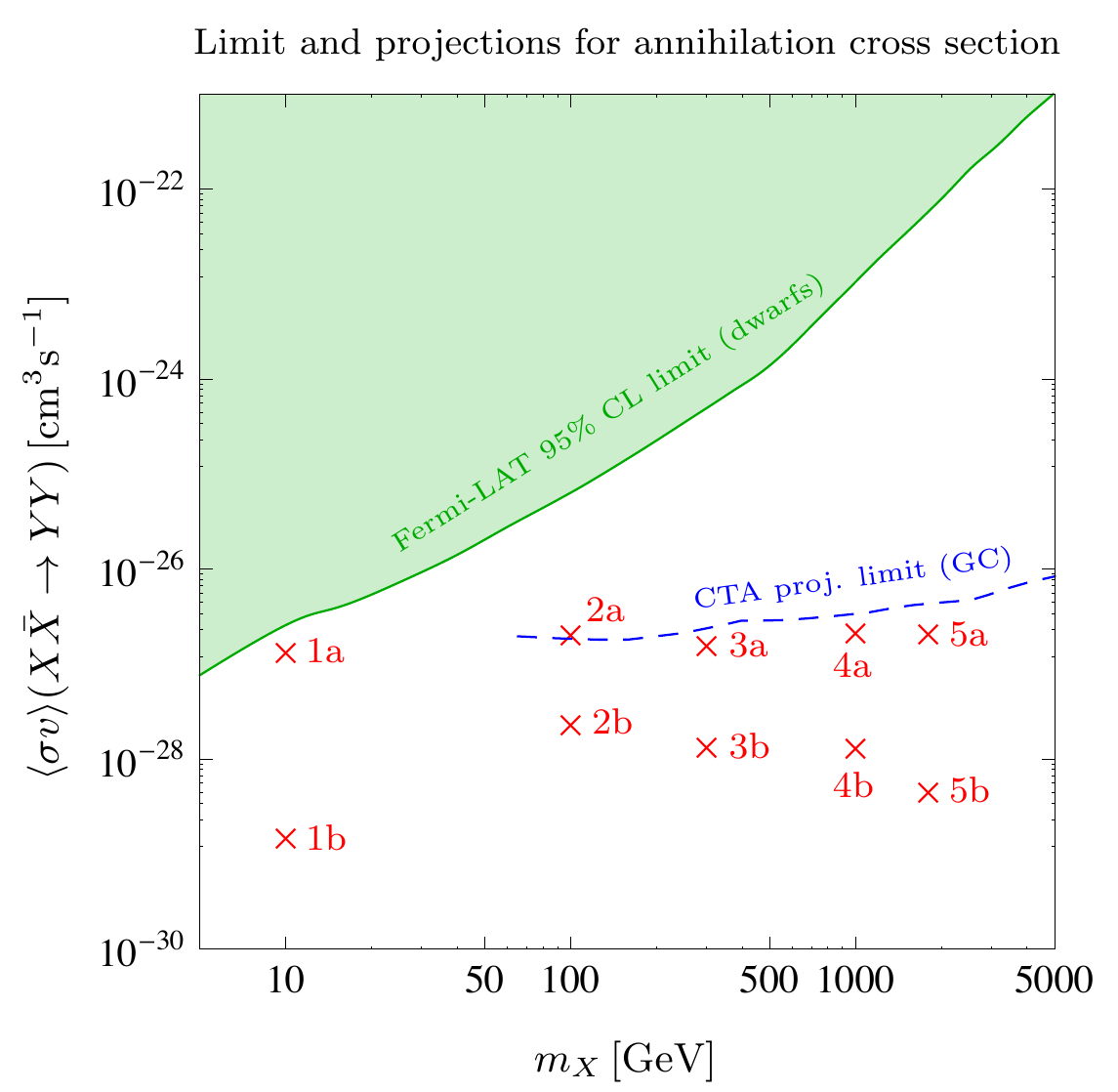}
\caption{Upper exclusion limit at 95\% C.L. on the dark matter annihilation cross section from the Fermi-LAT
observations of dwarf spheroidal galaxies (green curve and shaded area). The blue dashed curve denotes the 
projected exclusion sensitivity of CTA for the observations of the Galactic center (Einasto profile), taken from~\cite{Ibarra:2015tya}.
Red crosses denote our benchmark points.}
\label{fig:dwarflim}
\end{center}
\end{figure}

Dwarf spheroidal galaxies are among the most constraining environments for 
DM annihilating into $\gamma$-rays because of their large mass-to-light ratio. 
In the model (and parameter space) under consideration the dominant contribution 
to a continuum $\gamma$-ray flux is provided by the annihilation process 
$X\bar{X} \to YY \to 4 \gamma$. We have computed the corresponding bound 
from dwarf spheroidal galaxies using the Fermi-LAT public likelihood 
data~\cite{Fermi-LAT:2016uux} including the nine brightest dwarfs, which 
have been confirmed. The resulting 95\% C.L. upper limit on 
$\langle\sigma v\rangle(X\bar{X}\rightarrow YY)$ are shown in Fig.~\ref{fig:dwarflim}. 
They range from $\sim2 \times 10^{-27} \rm\, cm^3 s^{-1}$ for $m_X=10\,$GeV to 
$\sim 10^{-23} \, \rm cm^3 s^{-1}$ for $m_X=1\,$TeV.
While the limit is similar to the one for \emph{e.g.}~annihilation into $b\bar b$
for $m_X \lesssim 100\,$GeV~\cite{Fermi-LAT:2016uux} the limit becomes significantly weaker for high masses.
We show the constraint for $m_Y=50\,$MeV, however, the result is virtually insensitive to the mediator mass
as long as $m_Y < m_\pi \ll m_X$. For comparison,
 in Fig.~\ref{fig:dwarflim} we also display the
predicted annihilation cross section $\langle\sigma v\rangle(X\bar{X}\rightarrow YY)$ for the
considered benchmark 
points. While benchmark point 1a lies very close to the current sensitivity, for large masses, $m_X\gtrsim1\,$TeV, the cross sections are more than three (and up to six) orders
of magnitude below the limit.

Another sensitive target for DM annihilations is the Galactic center. Searches for a continuum $\gamma$-ray signal reach a sensitivity that is similar or slightly weaker than the one from dwarfs (depending on the considered DM density profile)~\cite{TheFermi-LAT:2017vmf}.

Besides the process $X\bar{X} \to YY \to 4 \gamma$, our model also leads to DM annihilation  via a mediator in the $s$-channel.
The processes $X\bar{X} \to t\bar{t}, \,b\bar b$ provide another contribution to the continuous spectrum of photons through
the hadronization and decay of the quarks.
Furthermore, the loop-induced annihilation process $X \bar{X} \to  \gamma \gamma$
provides the signature of monochromatic $\gamma$-ray lines (see \emph{e.g.}~\cite{Hektor:2017ftg}).
Both channels exhibit a dominant $s$-wave contribution to annihilation for $\sin\alpha \simeq 1$, \emph{i.e.}~for
a (mostly) pseudoscalar mediator.
However, as previously pointed out, we are primarily interested in
the part of the parameter space featuring small mediator masses and decay widths to enable the mediator to escape the Sun and produce observable solar $\gamma$-ray signals. In this region the $s$-channel annihilation
process is heavily suppressed by the smallness of the coupling $g_q$ and by the off-shellness of the mediator. 
Hence, for the considered parameter space the cross sections are too small to
provide a sensitivity competitive to the one arising from the process $X\bar{X} \to YY$.

Several high-energy $\gamma$-ray telescopes are supposed to commence operation in the next 5 years. Most notably, the ground-based Cherenkov Telescope Array (CTA)~\cite{2011ExA....32..193A}, with the effective area of several ${\rm km}^2$, has been developed to measure $\gamma$-ray fluxes in the $\TeV$ range. Due to its design, CTA can operate only during moonless nights and is hence not an appropriate experiment for solar observations. However, projected CTA sensitivity to  $\gamma$-ray fluxes in the Galactic center still make it an excellent probe of DM induced $\gamma$-rays. Projections indicate that CTA will be able to probe models with $\sigmav \sim 10^{-27}\, {\rm cm}^3 \rm s^{-1}$ over a wide range of DM masses with 100 hours of observation time~\cite{Ibarra:2015tya}. 
Figure~\ref{fig:dwarflim} shows the projected CTA limits for an Einasto DM density profile and
the ``wide box'' spectrum considered in Ref.~\cite{Ibarra:2015tya}, which provide a good estimate for our simplified model. 
Notice that most of our benchmark points are below the CTA sensitivity after 100 hours of observation time.

As we will show in the following sections,  solar $\gamma$-ray searches will be able to probe regions of our simplified model parameter space to a degree competitive and complementary with the observations of the Galactic center and of dwarf spheroidal galaxies using the next generation of $\gamma$-ray telescopes. 

\subsection{Big Bang Nucleosynthesis}

Our model contains long-lived mediators and is hence subject to constraints from BBN~\cite{Kawasaki:2004yh,Jedamzik:2006xz}. In order for the mediator not to inject significant amounts of energy into the primordial plasma during the period of BBN, starting roughly 1 min after the big bang, we impose a conservative bound on the mediator lifetime of $\tau_Y \lesssim 1\, {\rm s}$. As the $Y$ width in our model depends only on $m_Y$ and $g_q$ the BBN limit on the mediator lifetime will impose limits on a combination of the two model parameters without affecting $m_X$ and $g_X$. Figure~\ref{fig:bbn} illustrates the parameter region compatible with BBN predictions. Couplings to quarks of $\gtrsim 10^{-5}$ are allowed by BBN for $m_Y \sim \mathcal{O}(100)$ MeV, while larger $g_q$ are necessary for lower $m_Y$ in order for the mediator lifetime not to be too long to affect BBN.

\subsection {Cosmic Microwave Background} The constraints from CMB measurements imply an upper limit on the amount of energy that $Y$, produced by DM annihilation, can inject via its decay without affecting the recombination epoch ($\langle v \rangle \sim 10^{-7}$). In terms of the DM annihilation cross section, the limit can be written as~\cite{Ade:2015xua,Bringmann:2016din}
\begin{equation}
	\sigmav \lesssim 8\times 10^{-25} \text{cm}^3 \text{s}^{-1} \left( \frac{{\rm Br}_{\gamma \gamma }}{0.1} \right)^{-1}\left( \frac{m_X}{100 \GeV}\right)\,, \label{eq:cmb}
\end{equation}
where ${\rm Br}_{\gamma \gamma}$ is the fraction of the final state particles which end in photons, including showering, hadronization and subsequent baryonic decays.\footnote{The right-hand side of the equation should be divided by a factor of 2 in case of Majorana DM\@.} Equation~\eqref{eq:cmb} suggests that TeV scale DM annihilating into $\gamma$-rays with $\sim 100\%$ efficiency is constrained by CMB only if the annihilation cross section is orders of magnitude higher at recombination than the thermal $\sigmav \sim 3\times 10^{-26} {\rm cm}^3 \rm s^{-1}$.

Several papers have recently pointed out that large low energy cross section enhancements due to bound state and Sommerfeld-like dynamics impose stringent constraints on scalar or vector mediators for $m_Y \sim \mathcal{O}(1\!-\!100)$ MeV, see {\it e.g.}~Refs.~\cite{Bernal:2015ova,Bringmann:2016din}. In the following we argue that such effects are not significant in our model. 

Let us consider the scalar part of the interaction first. Assuming a Yukawa potential the condition to allow at least $N_l$ bound states of angular momentum $l$ is
\begin{equation}
	N_l \leq \frac{m_X}{2l+1} \frac{\alpha_X}{m_Y}\,, \label{eq:nl}
\end{equation}
where $\alpha_X = g_X^2 {\rm cos}^2 \alpha / 4\pi$ in our model. Equation~\eqref{eq:nl} follows from the Bargmann-Schwinger limit \cite{bargmann,schwinger}, however note that an equivalent expression can be obtained via a variational calculation (\emph{i.e.}~requiring the expectation value of the Hamiltonian to be negative).  For $l=0$ and setting $N_l = 1$  the bound gives
\begin{equation}
	\alpha_X \gtrsim \frac{m_Y}{m_X},
\end{equation}
giving the condition for existence of at least one DM bound state. For $m_Y \sim 100 \MeV$ and $m_X \sim 100\! -\!1000 \GeV$, the condition then requires $\cos \alpha \gtrsim {\cal O}(0.01\!-\!0.1)$, assuming $g_X \sim 1$ needed for the relic density. Such large mixing angles are already in tension with the direct detection constraints, leading us to conclude that bound state formation is not a significant effect in this model. The fact that direct detection can impose stringent bounds on bound state formation and Sommerfeld boost factors has also been discussed in Refs.~\cite{Arina:2010wv,Bernal:2015ova}.

In addition, Sommerfeld-like enhancements ($S$) are large only in the regime of $\langle v \rangle \lesssim \alpha_X$, and saturate at $S \sim m_Y/ (2 m_X)$. For models with $m_Y \sim 100$ MeV, $m_X \sim 500$ GeV, mixings of ${\rm cos }\, \alpha\sim 10^{-3}$ and $g_X \sim 1$ this implies that the enhancement will be at most of the order of $\mathcal{O}(10)$ only in the region of $\langle v \rangle \lesssim 10^{-7}$ and negligible or close to unity for larger velocities. We hence conclude that the low energy enhancements due to the scalar exchanges do not amount to significant effects in our model. 

Treatment of the pseudoscalar part of the potential is significantly more difficult. Potentials induced by pseudoscalar exchanges include tensor spin correlations as well as dominant terms scaling like $1/r^3$ where $r$ is the distance to the center of mass of two DM particles. The potential hence depends on the spin configuration of the $X\bar{X}$ system, and can lead to either enhancement or suppression of the cross section.  More importantly the $1/r^3$ dependence of the potential inevitably leads to Schr\"{o}dinger equation solutions which are divergent as $r\rightarrow 0$. Recently, Ref.~\cite{1126-6708-2009-11-046} provided a treatment of such potentials by introducing a regularization cutoff $r_0$ in order to compute the size of the enhancement. The end result depends on $r_0^2$ implying a quadratically ultraviolet divergent theory with no clear prescription of how to determine the cutoff $r_0$. The authors however do point out that other than in cases where bound state resonances appear at threshold, the enhancement due to pseudoscalar exchanges is generically close to unity.

\subsection {LHC Dark Matter / Mediator Searches} The LHC provides no useful bounds on our simplified model. As we are interested in the $m_X \gg m_Y$ regime, there is no significant MET+$j,Z,H,...$ signal in the model. In addition,  mediators with $m_Y \lesssim 100 \GeV$  are well beyond the scope of LHC dijet or multitop searches. 

\subsection{Flavor Constraints -- Beam Dump Experiments} The  $m_Y \sim (10\!-\!100) \MeV$ regime we consider in this analysis implies that  the mediator can be directly produced in effective quark flavor changing neutral current processes (FNCN), as for instance $b \rightarrow s + Y$ and $s \rightarrow d + Y$. As pointed out in Ref.~\cite{Dolan:2014ska}, the experimental constraints on deviations from the SM prediction in flavor-violating meson decays translate to strong constraints on the couplings of the mediator to the  SM for $m_Y < 10$ GeV. 

\begin{figure}[t]
\begin{center}
 \includegraphics[width=1\columnwidth]{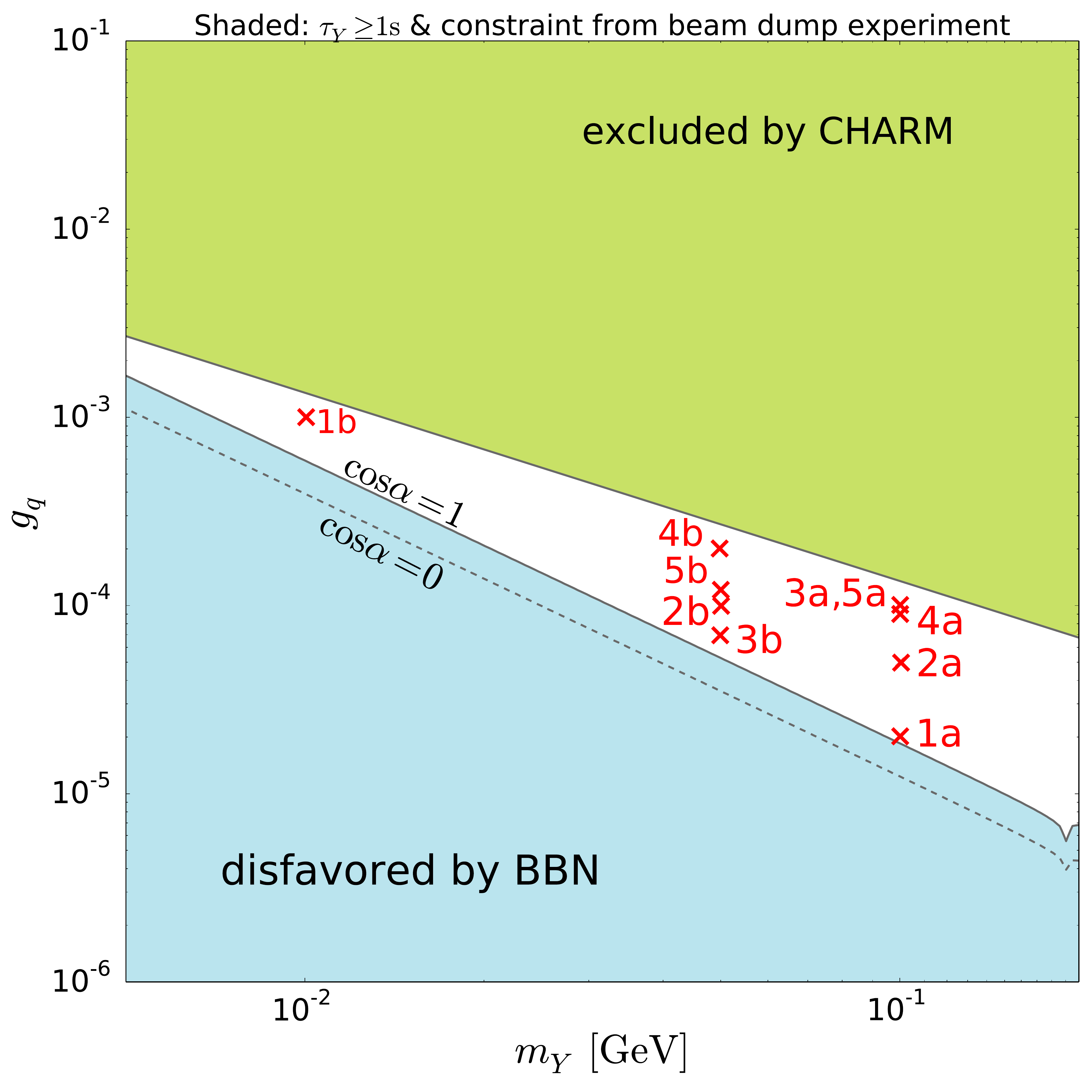}
 \caption{Constraints on long-lived mediators from BBN and beam dump experiments.  The shaded regions represent parts of the model parameter space disfavored by BBN (blue region) and flavor ruled out by the CHARM experiment at 90\% C.L. (green region). The benchmark points from Tab.~\ref{tab:bench} are shown as red crosses. Note that some points are too close to each other to be visually separable.}
\label{fig:bbn}
\end{center}
\end{figure}

From the flavor physics point of view, our scenario is phenomenologically similar to the one analyzed in Ref.~\cite{Dolan:2014ska}, since the contribution from scalar couplings with a Yukawa-like structure cancels in the divergent part of the effective FCNC vertex, leaving only the pseudoscalar component as the dominant (divergent) contribution. Interpreting the theory as the low energy limit of a renormalizable UV completion, we recover the results of Ref.~\cite{Dolan:2014ska} for their so-called ``quark Yukawa-like'' scenario, with the replacement $g_{Y_q} \rightarrow g_q \sin\alpha$. 

The orders of magnitude of $g_q$ we consider here are constrained only by beam dump experiments. Despite $g_q \ll 1$, light mediators can be copiously produced in such experiments, due to the large number of parent particles that can produce a mediator as a decay product. The CHARM experiment results~\cite{Bergsma:1985qz} provide the most stringent constraint in the considered region of parameter space~\cite{Bezrukov:2009yw,Essig:2010gu,Clarke:2013aya,Dolan:2014ska}. 

The CHARM Collaboration aimed at searching for axion-like particles directly produced from the interaction of protons on target. For the values of couplings considered in the present analysis, direct production of mediators is negligible~\cite{Dolan:2014ska}. However, a significant number of mediators can be produced in the decay on flight of $K$ and $B$ mesons.

In order to compute the number of expected events $N_{\rm events}$ (defined as the number of mediators decaying inside the detector), we consider the full two-body kinematics of the decay, instead of relying on analyses that have been performed using direct production.  The solid angle covered by the detector in the decaying meson rest frame depends on only two variables: the boost of the mother particle in the laboratory frame and the distance between the decay point and the detector itself. We approximate the detector to have cylindrical symmetry, with a length of 35 m and a transverse section $A=(6\times 4.8)$ m${}^2$, equal to the active area of the first (and larger) scintillation counter hodoscope in the CHARM detector. We also assume the decaying mesons to travel on the central axis of the detector, {\it i.e.}~the flux to be ideally focused. These approximations give an active volume larger than the actual setup, resulting in an overestimate of the number of expected events; \emph{a posteriori} we nonetheless find them reasonable given the resulting bound. After having defined the kinematics of the process, we compute $N_{\rm events}$ and require $N_{\rm events}<2.3$, \emph{i.e.}~the yield is compatible at 90\% C.L. with the nonobservation of a signal. Notice that $N_{\rm events}$ depends exponentially on the lifetime of the mediator in the laboratory frame, which is in turn proportional to its energy. A proper analysis, including the full simulation of the experiment in order to compute the energy spectrum of the produced particles, is beyond the scope of this work. Here we estimate the number of events assuming the pion energy spectrum computed in Ref.~\cite{Bergsma:1985qz}. 

Our results are summarized in Fig.~\ref{fig:bbn}. Note that the CHARM bound (green shaded region) on $g_q$ is poorly sensitive to the value of $\cos\alpha$ (for $\cos\alpha \ll 1$), hence we display it  for the pure pseudoscalar limit. The CHARM derived bound is relaxed by almost a factor of 2 with respect to the one derived in~\cite{Dolan:2014ska}. In the mediator mass range $(5\!-\!100) \MeV$, the CHARM bound is well  approximated by
\begin{equation}
g_q  \lesssim 1.35 \times 10^{-5} \left(\frac{m_Y}{\rm GeV} \right)^{-1}\,. 
\end{equation}

Figure~\ref{fig:bbn} reveals an interesting interplay between the BBN requirement on the mediator lifetime, which constrains the size of $g_q$ from below (blue shaded region) and the flavor bound, which on the contrary constrains $g_q$ from above. The two requirements leave only a narrow band (white region) of allowed values for the mediator-quark couplings, right in the ballpark to produce sizable solar $\gamma$-ray signals.

Figure~\ref{fig:bbn} also illustrates the important role that beam dump experiments could play in further constraining DM models with long-lived mediators.

\section{Solar $\gamma$-rays}\label{sec:solar}

\subsection{Solar Capture Rate}

Assuming that the dominant DM nucleon scattering originates from the spin-independent interaction in Eq.~\eqref{eq:dd}, the solar capture rate can be written as \cite{Jungman:1995df}
\begin{widetext}
\begin{equation}
	C_\text{cap} = 4.8 \times 10^{24} {\rm s}^{-1} \left( \frac{\rho_\odot} { 0.3 \GeV {\rm cm^{-3}}} \right) \left( \frac{\text{GeV}}{m_X }\right) \left( \frac{270\, {\rm km} \rm s^{-1}}{\bar{v} }\right)
				\sum_i 
			F_i(m_X) \left( \frac{\sigma_{XN_i}}{10^{-40} {\rm cm}^2}\right) \times f_i\times  \phi_i \times S\left(\frac{m_X}{m_{N_i}}\right)\left( \frac{\text{GeV}}{m_{N_i}} \right)\,, \label{eq:cap}
\end{equation}
\end{widetext}
where $\rho_\odot$ is the local halo density of DM, $\bar{v}$ is the average local velocity of DM, $F_i$ are the suppression form factors for individual nuclei species which make up the Sun and $f_i$ are the mass fractions of the $i^{\rm th}$ element. The coefficients $\phi_i$ represent the densities of individual elements in the Sun, while $S$ is the kinematic suppression factor. More detail on the definition and values of the quantities in Eq.~\eqref{eq:cap} can be found in Ref.~\cite{Jungman:1995df}. It is important to mention that since the spin-independent DM nucleus cross section scales as $A^2,$ where $A$ is the total number of protons and neutrons in the nucleus, the heavier elements in the Sun can provide significant contribution to the overall capture rate despite their lower abundance. This is contrary to the case of spin-dependent scattering, where typically only scattering off hydrogen is significant. 

The annihilation rate of DM in the center of the Sun is proportional to 
the density of DM particles:
\begin{equation}
\Gamma_{\rm ann } = \frac{N^2}{4 V_{\rm eff}}  \left[ \langle \sigma v \rangle (X\bar{X} \rightarrow YY)+\langle \sigma v \rangle (X\bar{X} \rightarrow t\bar{t}) \right]   \label{eq:annrate}\,,
\end{equation}
where $N$ is the number of captured DM particles and $V_{\rm eff}= 5.8\times10^{30}{\rm cm}^3 (\GeV / m_X)^{3/2} $ is the effective volume of the Sun~\cite{Griest:1986yu,Gould:1987ir,Halzen:2005ar,Andreas:2009hj}. The expression is valid for Dirac fermionic DM candidates and should be multiplied by a factor of 2 on the right-hand side if DM is composed of self-conjugate particles. 

The total number of DM particles in the Sun is a result of the competing capture and annihilation processes and is described by the differential equation~\cite{Jungman:1995df}
\begin{equation}
\frac{{\rm d}N}{{\rm d}t} = C_\text{cap} - 
C_\text{ann} N^2 
\,, \label{eq:Riccati}
\end{equation}
where $C_\text{ann}=2\Gamma_{\rm ann }/N^2$ is independent of $N$. 
Assuming that the Sun has been accumulating DM during its entire lifetime we can solve Eq.~\eqref{eq:Riccati} for $t=t_\odot \simeq1.5 \times 10^{17} \,\text{s}$, obtaining an expression for the number of DM particles today:
\begin{equation} 
 N = \sqrt{\frac{C_\text{cap}}{C_\text{ann}}} \tanh\left(\sqrt{C_\text{cap} C_\text{ann}} \,t_\odot\right)\,. \label{eq:solN}
\end{equation}
From $N$ we obtain the annihilation rate via Eq.~\eqref{eq:annrate}.

From sufficiently large rates
\begin{equation}
	\sqrt{C_\text{cap} C_\text{ann}} \,t_\odot \gg1\,, \label{eq:eqcond}
\end{equation}
which implies that equilibrium between capture and annihilation is reached, \emph{i.e.}~${\rm d}N/{\rm d}t = 0$. 
In equilibrium $\Gamma_{\rm ann}=1/2\, C_\text{cap}$ is not sensitive to the 
DM annihilation cross section anymore but allows one to extract information on the DM-nucleon scattering cross section. For a nonequilibrium scenario, on the other hand,
$\Gamma_{\rm ann}$ contains information about both $C_\text{cap}$ and $C_\text{ann}$,
which could potentially be exploited to determine the annihilation cross section provided an independent measurement of the DM-nucleon scattering cross section 
from direct detection experiments.

For most of the considered benchmarks the left-hand side of Eq.~\eqref{eq:eqcond} is (considerably) smaller than
unity, \emph{i.e.}~the equilibrium condition is not satisfied, as shown in Tab.~\ref{tab:fluxes}. In fact, the spin-independent DM-nucleon scattering cross section allowed by LUX is generically not large enough to provide equilibrium for an annihilation cross section around $3 \times 10^{-26} \rm cm^3 s^{-1}$ or lower.

In the following section, we will show that despite the fact that equilibrium is difficult to reach, the simplified model we consider can result in solar $\gamma$-ray fluxes large enough to be observed by the next generation of $\gamma$-ray observatories. 

\subsection{$\gamma$-Ray Spectral Shape}\label{sec:kinematics}

Given that the escape velocity of the Sun is approximately $10^{-3}$, the captured DM particles are nonrelativistic,  
resulting in kinematics well approximated by two DM particles annihilating at rest. In the lab frame, the final state consists of two anticollinear mediators, each of energy $m_X$, which subsequently decay into two photons each.

\begin{figure*}[t]
\begin{center}
 \includegraphics[width=0.45\textwidth]{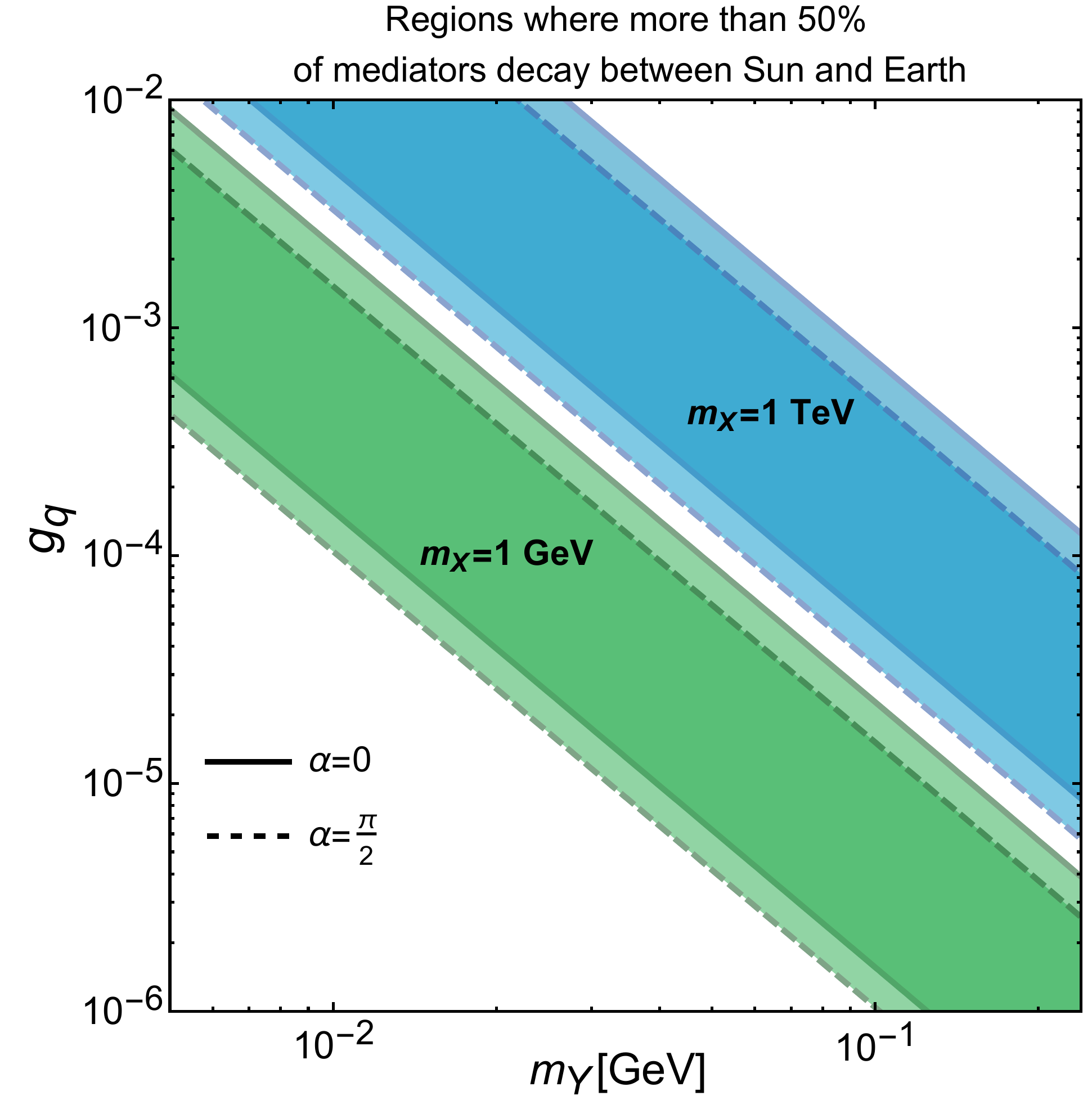}\,\,\,\,\,\,\,\,\,\,\,\,\,
  \includegraphics[width=0.45\textwidth]{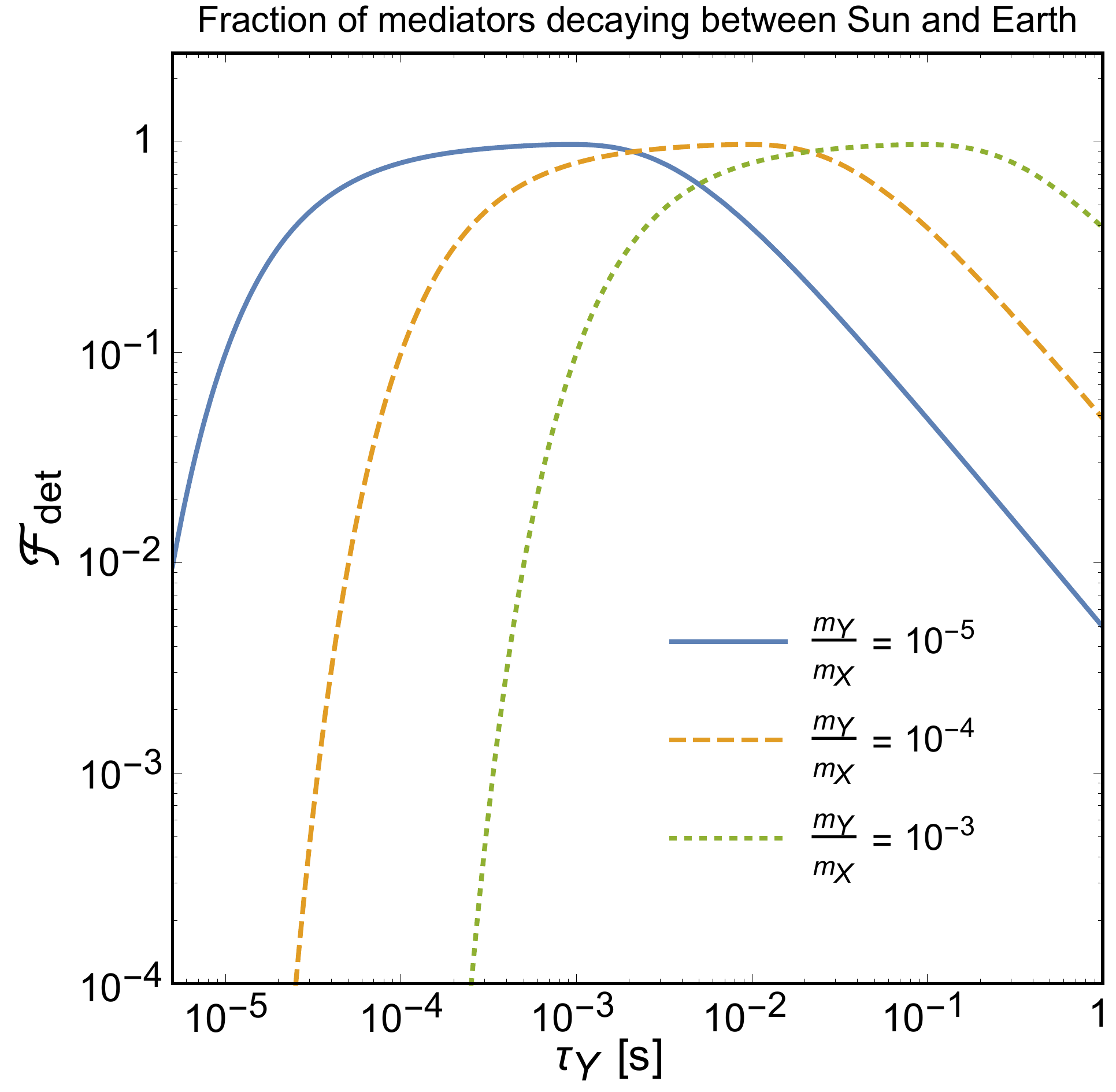}
\caption{Left: The shaded regions indicate the parameter space which satisfies the requirement that at least 50\% of the mediators produced in the Sun decay between the solar surface and the Earth for two DM masses, as labeled. The solid/dashed boundaries are limits of the pure scalar/pseudoscalar mediators. Right: Fraction of mediators decaying between the solar radius and the Earth's orbit, as a function of the mediator lifetime and for different values of mediator boost, as labeled. }
\label{fig:decayok}
\end{center}
\end{figure*}

Given the axis specified by the momentum of the mediator in the lab frame, we define $\theta^*_i$ as the angle between one emitted photon and that axis, as measured in the mediator's rest frame. The photon energy in the lab frame is then
\begin{equation}
E^\gamma_i = \frac{m_X}{2} \left(1+\cos\theta^*_i \sqrt{1-\frac{m_Y^2}{m_X^2}}\right)\,.
\end{equation}
Notice that the two emitted photons have $\cos\theta^*_2 = -\cos \theta^*_1$, and thus $E^\gamma_1+ E^\gamma_2 = m_X$. Since the mediator is a spinless particle, the decay is isotropic in its rest frame, implying that the differential decay width is flat in the variable $\cos\theta^*$. At sufficiently large distances, such that only one of the two photons is detected, the resulting energy spectrum for the photons is a flat box~\cite{Ibarra:2012dw}:
\begin{equation}
\label{eq:inf_phflux}
\frac{{\rmd} N_\gamma}{{\rmd} E_\gamma } = N_Y \frac{2 }{\Delta E} \Theta\left(E_\gamma-E_-\right)\Theta\left(E_+-E_\gamma\right),
\end{equation}
where $ N_Y$ is the number of mediators giving rise to detectable photons,  the factor 2 takes into account that two photons are produced per mediator decay, and
\begin{eqnarray}
E_{\pm} &=& \frac{m_X}{2} \left(1\pm \sqrt{1-\frac{m_Y^2}{m_X^2}}\right)\,,\non 
\Delta E &=& m_X \sqrt{1-\frac{m_Y^2}{m_X^2}}\,.
\end{eqnarray}
The quantity $N_Y$ is generally smaller than the number of mediators produced in DM annihilation since, depending on the considered celestial body, some decays could give rise to nondetectable photons. In our setup, a mediator produced in the center of the Sun only gives rise to detectable photons if it decays between the solar radius $R_\odot$ and the Earth's orbit $D_\odot$. The fraction of mediators decaying in this region is
\begin{equation}
\label{eq:fracdet}
{\cal F}_\text{det} =  \left(e^{-\frac{R_\odot}{\gamma \beta c \tau_Y}}-e^{-\frac{D_\odot}{\gamma \beta c \tau_Y}}\right),
\end{equation}
where $\gamma$ and $\beta$ are the boost and velocity of the mediator ($\gamma \beta = \sqrt{m_X^2/m_Y^2-1}\simeq m_X/m_Y$) and $\tau_Y$ is its lifetime. We report the value of ${\cal F}_\text{det}$ for the different benchmark points in the third column of Tab.~\ref{tab:decay}. Most of the points feature at least $\mathcal{O}(50\%)$ of decays in the required region, except for benchmark point 5b where the boosted mediators tend to live too long and decay mostly beyond the Earth's orbit. Figure~\ref{fig:decayok} left illustrates the simplified model parameter space leading to a large fraction of mediators decaying between the Sun and the Earth. A large range of $g_q$ values satisfies the requirement that at least $50\%$ of mediators decay within 1 astronomical unit of the Sun. A comparison to Fig.~\ref{fig:bbn} reveals that a substantial portion of this parameter space is also  compatible with the BBN and flavor bounds. The dependence on the mixing angle $\cos\alpha$ is very mild, as anticipated. Figure~\ref{fig:decayok} right reports the quantity ${\cal F}_\text{det}$ as a function of the mediator lifetime, for different choices of the mediator boost: too short-lived mediators will mostly decay inside the Sun, while too long-lived mediators will decay beyond the Earth's orbit. It is evident from the figure that no fine-tuning of $\tau_Y$ is required, since the lifetime range where a sizable ($\gtrsim 10\% $) fraction of mediators decay into detectable photons spans multiple orders of magnitude.

At finite distances Eq.~\eqref{eq:inf_phflux} needs to be corrected in order to account for the possibility of correlated events involving both of the produced photons. To compute these corrections we notice that the direction of a photon in the lab frame with respect to the mediator's momentum is given by the angle $\theta_i$, such that
\begin{equation}
\label{eq:cos_lab}
\cos \theta_i = \frac{\cos \theta^*_i+\sqrt{1-\frac{m_Y^2}{m_X^2}}}{\sqrt{\left(\cos \theta^*_i+\sqrt{1-\frac{m_Y^2}{m_X^2}}\right)^2+\frac{m_Y^2}{m_X^2} \left(1-\cos^2\theta_i^*\right)}}\,.
\end{equation}
For values of $m_Y/m_X\ll 1$, $\cos \theta_i \simeq 1$ for almost any value of $\theta^*_i$, except in a small region around $\cos \theta^*_i = -1$ where $\cos \theta_i \simeq -1$. Given the values of $m_Y/m_X$ explored in our benchmark points ($m_Y/m_X<10^{-3}$) we can safely assume that both photons are emitted collinearly with the mediator momentum. The observed signal will depend on the spatial separation of the two photons at the target, leading to detection of either one or both photons.

It is hence important to address the question of whether Earth-based (or near-Earth-based) $\gamma$-ray observatories will be able to detect one or both photons from the $Y$ decay. The distance of a single photon from the mediator direction is given by $\Delta r = d\times \tan\theta_i$, where $d$ is the distance between the detector and the decay point. The spatial separation between the two photons, after they have traveled a distance $d$, will be
\begin{equation}
\label{eq:spat_sep}
\Delta r = 2d \frac{ \sqrt{\left({\cos\theta^*}^2-1\right) \left(\frac{m_Y^2}{m_X^2}-1\right) \frac{m_Y^2}{m_X^2}}}{1-{\cos\theta^*}^2-\frac{m_Y^2}{m_X^2}}
\end{equation}
for $\left|\cos\theta^*\right| < \sqrt{1-m_Y^2/m_X^2}$, while for $\left|\cos\theta^*\right| > \sqrt{1-m_Y^2/m_X^2}$ one of the photons will be emitted backwards with respect to the mediator momentum. %
\begin{table}[t]
{ \renewcommand{\arraystretch}{1.5}
\caption{Decay behavior of the mediators for the benchmark model points defined in Tab.~\ref{tab:bench}.  The third column denotes the fraction of mediators decaying between the solar surface and the Earth.  The column ${\cal F}_{2\gamma}$ displays the fraction of  events where both photons could be detected
 in satellite and terrestrial $\gamma$-ray observatories for DM masses below and above 300 GeV, respectively, following Eq.~\eqref{eq:lineratio}.  \label{tab:decay} }
\begin{tabular}{ccccc}
Benchmark  & $m_Y/m_X$ & $\tau\,[\text{s}]$ &  ${\cal F}_\text{det}$  & ${\cal F}_{2\gamma}$  \\
\hline\hline
1a &  $0.01$ & $0.19$ & $0.88$ & $4.3 \times 10^{-20}$ \\
1b &  $0.001$ & $0.076$ & $0.97$ & $3.3 \times 10^{-11}$ \\
2a &  $0.001$ & $0.031$ & $0.93$ & $4.7 \times 10^{-15}$ \\
2b &  $0.0005$ & $0.061$ & $0.96$ & $4.8 \times 10^{-10}$ \\
3a &  $0.00033$ & $0.0076$ & $0.9$ & $8.3 \times 10^{-17}$ \\
3b &  $0.00017$ & $0.12$ & $0.48$ & $1.4 \times 10^{-8}$ \\
4a &  $0.0001$ & $0.0094$ & $0.97$ & $9.1 \times 10^{-7}$ \\
4b &  $5 \times 10^{-5}$ & $0.015$ & $0.8$ & $2.7 \times 10^{-5}$ \\
5a &  $5.6 \times 10^{-5}$ & $0.0076$ & $0.96$ & $6 \times 10^{-6}$ \\
5b &  $2.8 \times 10^{-5}$ & $0.042$ & $0.28$ & $0.0001$ \\
\hline
\end{tabular} }
\end{table}

Equation~(\ref{eq:spat_sep}) has a minimum for $\cos\theta^*=0$, for which
\begin{equation}
\label{eq:r_min}
\Delta r_\textrm{min} = \frac{2d}{\sqrt{\frac{m_X^2}{m_Y^2}-1}}\simeq 2 d \frac{m_Y}{m_X} .
\end{equation}
For mediators emitted at the surface of the Sun the minimal separation between the two photons at Earth is \emph{e.g.}~$2.9\times 10^{4}\, {\rm km}$, for $m_Y/m_X=10^{-4}$, using $d=D_\odot=1\,  \rm AU = 1.496\times 10^8 $ km.

If $\Delta r_\textrm{min}$ is smaller than the typical length $\ell$ of one side of a (hypothetical) square detector ($\ell\sim 1$ m for Fermi-LAT/HERD satellites and $\ell\sim 10^3$ m for the HAWC/LHAASO detectors) then it  is in principle possible to detect both photons, giving a peculiar signal of two simultaneous events with total energy $m_X$. This condition translates into a requirement on the distance between the detector and the decay point
\begin{equation}
\label{eq:2ph_cond}
d < \frac{1}{2} \frac{m_X }{m_Y}\ell \,.
\end{equation}
The fraction of events that can potentially give a two-photon signal can be estimated as
\begin{equation}\label{eq:lineratio}
{\cal F}_{2\gamma}<  \frac{e^{-\frac{D_\odot-d}{\gamma \beta c \tau}}-e^{-\frac{D_\odot}{\gamma \beta c \tau}}}{{\cal F}_\text{det}}\,,
\end{equation}
where $d$ depends on the detector size through Eq.~\eqref{eq:2ph_cond} and the inequality results because Eq.~\eqref{eq:r_min} gives the minimal possible spatial separation. The fraction of $2\gamma$ events expected in satellites or ground-based telescopes for the benchmark model points is given in Tab.~\ref{tab:decay} (fourth column) and turns out to be generically $\ll 1$. The result suggests that the signal of DM annihilation in the Sun in our simplified model would in most cases be the observation of a continuum $\gamma$-ray signal with an energy cutoff of $\sim m_X$ (without an expected correlated signal of two-photon events). 

\subsection{Solar $\gamma$-ray Flux}
\begin{table}[t]
{ \renewcommand{\arraystretch}{1.5}
\caption{Solar $\gamma$-ray fluxes for the benchmark model points defined in Tab.~\ref{tab:bench}.  The column labeled $\Phi_\gamma^{\odot}$ denotes the total integrated flux, while the column titled $N^{\odot}_\gamma$ shows the naive estimate of a number of events which we can expect to observe using a detector with an area of $60\times 60\, {\rm cm}^2$ 
for masses below 300 GeV, and an Earth-based detector with $20 \times 10^3\,  {\rm m^2}$ effective area for masses above 300 GeV. In both cases we assume 1-year exposure. 
The last column shows the numerical value on the left-hand side of the equilibrium condition in Eq.~\eqref{eq:eqcond}. \label{tab:fluxes} }
\begin{tabular}{cccccc}
Benchmark & $\Phi_\gamma^{\odot} [{\rm cm}^{-2} \,\rm s^{-1}]$ & $N_\gamma^\odot (1\, {\rm yr})$  
 & $\sqrt{C_{\rm cap} C_{\rm ann}} t_\odot$ \\
\hline\hline
1a &   $1.6 \times 10^{-15}$ & $0.00018$ & $0.0039$ \\
1b &  $1.1 \times 10^{-10}$ & $12$ & $0.02$ \\
2a &   $7 \times 10^{-11}$ & $8.0$ & $0.15$ \\
2b & $5.2 \times 10^{-12}$ & $0.59$ & $0.044$ \\
3a &  $5.7 \times 10^{-10}$ & $64$ & $0.36$ \\
3b &  $2.1 \times 10^{-12}$ & $0.24$ & $0.065$ \\
4a & $1.3 \times 10^{-9}$ & $8.3\times 10^{9}$ & $0.79$ \\
4b &  $2.2 \times 10^{-11}$ & $1.4\times 10^{8}$ & $0.14$ \\
5a &  $1.4 \times 10^{-9}$ & $1.6 \times 10^{10}$ & $1.1$ \\
5b &  $9 \times 10^{-13}$ & $1\times 10^{8}$ & $0.1$ \\
\hline
\end{tabular} }
\end{table}

The simplified model we consider leads to solar $\gamma$-ray fluxes, characterized by photons with box-shaped energy spectra in the range of $m_Y$ to $m_X$. The flux we expect on Earth, assuming that DM annihilation takes place in the center of the Sun, is given by
\begin{eqnarray}\label{eq:diffflux}
\frac{\de \Phi_\gamma^\odot}{\de E} & = &  \frac{\langle \sigma v \rangle (X\bar{X} \to YY)}{\langle \sigma v \rangle (X\bar{X} \to \text{all})} \frac{\Gamma_{\rm ann}}{4 \pi D_\odot^2} 
{\cal F}_\text{det} \nonumber\\
 & & \times\frac{4}{\Delta E} \Theta\left(E_\gamma-E_-\right)\Theta\left(E_+-E_\gamma\right)\,.
\end{eqnarray}
For the considered benchmark points the ratio $\langle \sigma v \rangle (X\bar{X} \to YY)/\langle \sigma v \rangle (X\bar{X} \to \text{all})$ is always close to unity, as discussed at the end of Sec.~\ref{sec:sigann}. 

Table~\ref{tab:fluxes} shows the magnitude of the solar $\gamma$-ray flux  for the benchmark points defined in Tab.~\ref{tab:bench}, where in each case we assumed that the detector is 1 astronomical unit away from the Sun. Our results suggest that it is possible to expect fluxes as large as ${\cal O}(10^{-12} \!- \!10^{-9})\, {\rm cm}^{-2} \rm s^{-1}$ in our simplified model, while being allowed by other experimental constraints. For the considered benchmark points with  $m_X \sim 1 \TeV$ the solar fluxes are larger by factors of  ${\cal O}(50)$ and ${\cal O}(10^5)$ compared to the corresponding
total fluxes expected from the observation of the Galactic center and the brightest dwarf spheroidal galaxies,
respectively.

The ability of near future experiments to observe solar $\gamma$-ray fluxes from Tab.~\ref{tab:fluxes} depends in part on the levels of $\gamma$-ray backgrounds we expect from the Sun.

Solar models and solar observations  predict that the Sun is a poor source of $\gamma$-rays with energies above ${\cal O}(\text{GeV})$. References~\cite{2011ApJ...734..116A} and~\cite{Ng:2015gya}, based on Fermi-LAT 1.5- and 6-year data, provide a measurement of the solar $\gamma$-ray flux. The measured total flux level is of the level of $\sim 1.3\times 10^{-8} {\rm cm^{-2}} \rm s^{-1}$ for $E_\gamma \sim 10 \GeV$. The flux at the largest energy observed is $\sim10^{-10} {\rm cm^{-2}} \rm s^{-1}$ at $E_\gamma \sim 100 \GeV$. The authors of Ref.~\cite{Seckel:1991ffa} argued that such $\gamma$-rays are produced by high-energy cosmic rays scattering off solar photons and  the solar atmosphere, but their initial estimate underestimates the measured photon flux by one order of magnitude. While there are no measurements of solar $\gamma$-rays $\gtrsim 100 \GeV$, the authors of Ref.~\cite{Zhou:2016ljf} provide a model for the $\gamma$-ray flux which one would expect to observe from high-energy cosmic rays scattering off the solar outskirts. The authors predict the upper limit on the flux of $\sim \TeV$ scale $\gamma$-rays of $\sim 10^{-13} {\rm cm^{-2}} \rm s^{-1}$. It is important to note that the high-energy solar $\gamma$-ray emissions originating from high-energy cosmic rays should be spatially localized away from the center of the Sun, because the effects of the magnetic field can be neglected. As we will discuss shortly, this feature is important in the case of satellites like Fermi-LAT or HERD, which feature very good angular resolution and could hence (in principle) distinguish these emissions from the $\gamma$-ray signals in the center of the solar disk. Even though, in principle, it would be possible to spatially separate the signals from DM annihilation from the emissions originating from cosmic-rays, in discussing the prospects for detection in the next section we proceed under the following assumptions: (i) for $m_X \lesssim 100$ GeV we consider the observed solar $\gamma$-ray flux as the irreducible background and (ii) for $m_X \gtrsim 100$ GeV since the high-energy solar $\gamma$-ray background is well below the experimental sensitivities, we consider that these latter set the actual expected sensitivity to the model.

\subsection {Future prospects}

In the following  we provide a discussion of the ability of the Fermi-LAT~\cite{Atwood:2013rka}, HERD~\cite{Huang:2015fca}, HAWC~\cite{Abeysekara:2013tza} and LHAASO~\cite{Zhen:2014zpa,He:2016del} experiments to detect solar $\gamma$-rays originating from DM annihilation in the Sun.

To assess the Fermi-LAT sensitivity curve for the full time mission (10 years) in the vicinity of the Sun we consider the differential sensitivity for the position (0,90) in Galactic coordinates~\cite{fermi}, which ranges from $5 \times 10^{-9}\,  {\rm GeV\,  cm}^{-2} \rm s^{-1}$ for $E_\gamma \sim 100 \GeV$ to $ \sim 3 \times 10^{-8}\,  {\rm GeV \, cm}^{-2} \rm s^{-1}$ for $E_\gamma \sim 1 \TeV$. The electromagnetic calorimeter on board has an area of roughly $60\times 60\, {\rm cm}^2$, while the tracker offers excellent angular resolution of $\sim 0.15$ degrees. The angular resolution of Fermi-LAT is important as it will allow the detector to efficiently resolve the Sun in the sky (the Sun appears roughly 0.6 degrees in size at the distance of 1 astronomical unit). In turn, the very good angular resolution may allow Fermi-LAT to veto backgrounds originating from high-energy cosmic-ray scattering off the solar outskirts.

A similar reasoning holds as well for the HERD cosmic-ray detector to be deployed on the Chinese space station circa 2020, which is expected to have even better angular resolution ($\sim 0.1$ degrees) than the Fermi-LAT satellite with a similar effective area. The projected sensitivity of HERD assuming 5 years of exposure 
is expected to exceed the Fermi-LAT sensitivity for the full time mission for energies above $\sim 300\,$GeV.\footnote{%
Xiaoyuan Huang, private communication.} Note that at present no estimate of the differential 
sensitivity for a continuum emission has been provided by the HERD Collaboration.

\begin{figure}[t]
\begin{center}
 \includegraphics[width=1.\columnwidth]{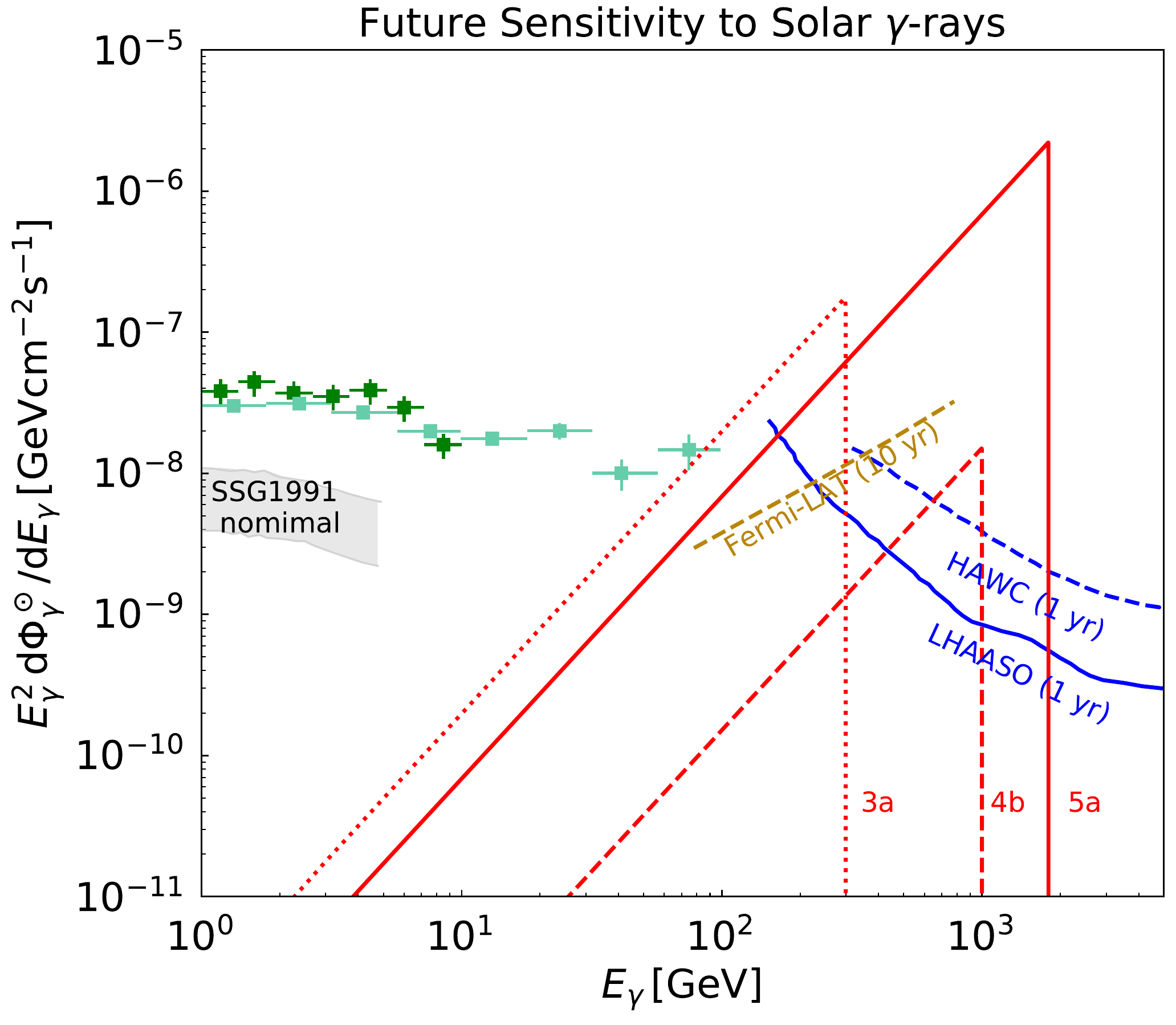}
\caption{Sensitivity of the Fermi-LAT~\cite{fermi}, HAWC~\cite{Abeysekara:2013tza} and LHAASO~\cite{Zhen:2014zpa,He:2016del} experiments as labeled, to probe the simplified model from Sec.~\ref{sec:mod} via solar $\gamma$-ray observations. A few benchmark point spectra from Tab.~\ref{tab:bench} are shown with red lines, as labeled. The observed solar $\gamma$-ray flux is depicted by the green data points, from Fermi-LAT 1.5-year data~\cite{2011ApJ...734..116A} (dark green) and from the analysis of Ref.~\cite{Ng:2015gya} of the Fermi-LAT 6-year data (light green). The gray band shows the flux magnitude predicted by Seckel \emph{et al.}~\cite{Seckel:1991ffa}.} 
\label{fig:id}
\end{center}
\end{figure}
\begin{figure}[!]
\begin{center}
 \includegraphics[width=1.\columnwidth]{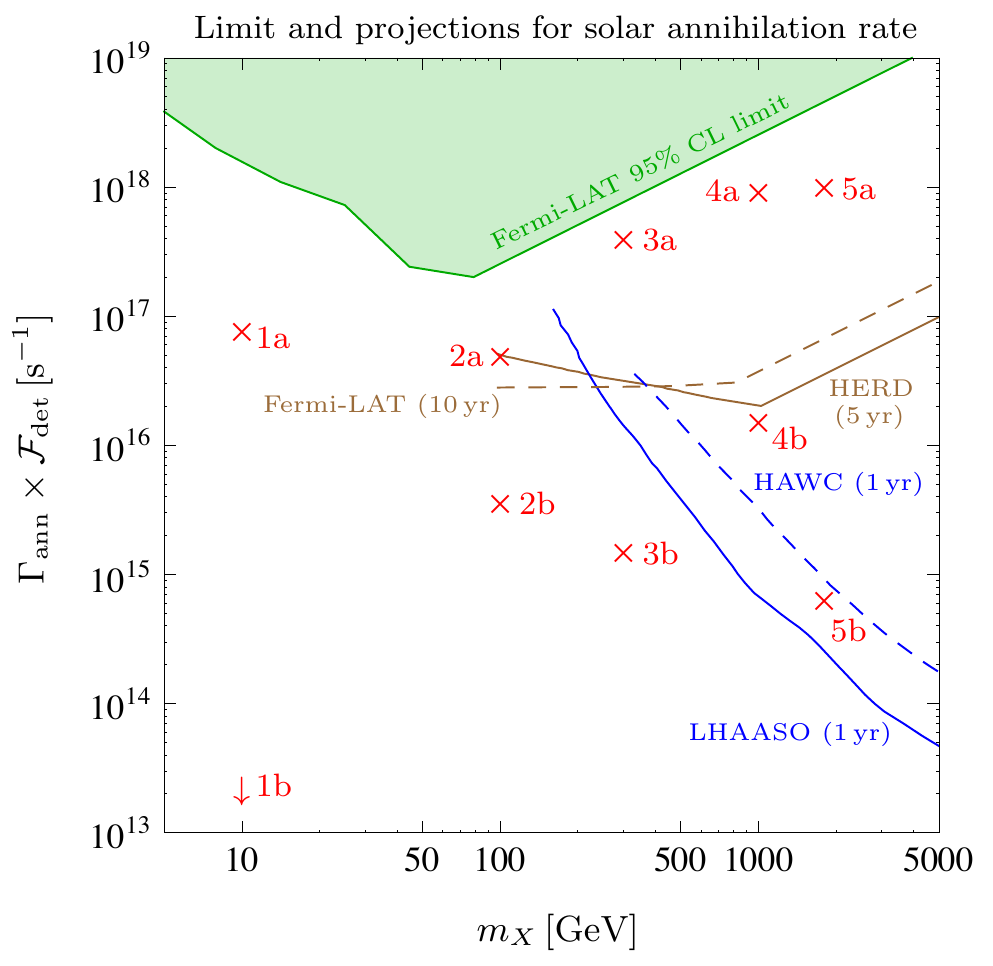}
\caption{Existing and projected upper 95\% C.L. exclusion limits on the dark matter annihilation rate from the solar $\gamma$-ray observations, as a function of the dark matter mass. The shaded area is excluded by the existing Fermi-LAT measurements~\cite{Ng:2015gya}. The benchmark points from Tab.~\ref{tab:bench} are shown as red crosses.}
\label{fig:sunbckg}
\end{center}
\end{figure}

Another experiment which can perform solar $\gamma$-ray observations is LHAASO, a water Cherenkov detector array scheduled to commence operation in 2017. Due to poorer angular resolution, LHAASO will likely not be able to efficiently distinguish $\gamma$-rays induced by cosmic ray scattering events from DM annihilation induced $\gamma$-rays based on the direction of the incoming $\gamma$-ray, but will be able to resolve an area roughly the size of the Sun.  However, what LHAASO might lack in angular and energy resolution ($\sim 20\%$), it compensates for with a large effective area (depending on the energy of the $\gamma$-rays, the effective area varies from $\sim 10^4 \, \rm m^2$ for $E_\gamma >500$ GeV to $\sim 6 \times 10^4\,  \rm m^2$ for $E_\gamma>1$~TeV), making it overall a very sensitive high-energy $\gamma$-ray detector. 

Compared to HERD, LHAASO will certainly suffer from more background contamination both from the cosmic-ray induced solar $\gamma$-ray fluxes, and from the diffuse $\gamma$-ray background. LHAASO also features very good $\gamma$/proton discrimination power for  $E_{\gamma/p} \gtrsim 100 \GeV$, suggesting that the experiment should be able to significantly reduce the backgrounds stemming from cosmic-ray misidentification. Despite larger backgrounds, we find that LHAASO will offer an excellent probe of DM induced $\gamma$-ray fluxes originating from the Sun.  

Finally, a water Cherenkov detector experiment HAWC,  which is currently running with a lower effective area than LHAASO ($\sim 10^4 \rm \,m^2$ above 1 TeV~\cite{Abeysekara:2013tza}), already displays sufficient sensitivity to observe solar $\gamma$-ray fluxes $\lesssim 10^{-12}\, {\rm cm}^{-2} \rm s^{-1}$ for $E_\gamma > 1000$ GeV. 

Figure~\ref{fig:id} illustrates the experimental sensitivity together with the photon energy spectrum produced for three of the considered benchmark points.\footnote{We omit a smearing of the spectra according
to the detector resolution in Fig.~\ref{fig:id}.} 

Figure~\ref{fig:sunbckg} displays the 95\% C.L. exclusion limit on the DM annihilation rate derived under the assumption that the $\gamma$-ray differential flux in Eq.~\eqref{eq:diffflux} does not overcome the solar flux observed by Fermi-LAT by more than $2 \sigma$.
The shape of the exclusion limit can be understood as follows. Due to the box-shaped spectra ($\propto E_\gamma^2$ in the representation of Fig.~\ref{fig:id})
for masses above 75\,GeV the constraint arises from the data point with the largest energy. The resulting 
limit is, hence, proportional to $m_X$.\footnote{%
Note that in~\cite{Ng:2015gya} no information on the flux above 75\,GeV is provided. An upper limit on the flux in the higher energy bins would potentially strengthen the limit for large masses.
}
On the other hand, for DM masses below 75\,GeV information from the other bins are exploited as well. The most constraining bin is in fact the one close to the DM mass under investigation. Since the measured flux increases at low energy, the upper exclusion limit becomes less constraining for light DM. The bound is most constraining for $m_X \gtrsim 100$ GeV, limiting the value of $g_q$ even further than the CHARM bound (\emph{c.f.}~Fig.~\ref{fig:bbn}).

Additionally, Fig.~\ref{fig:sunbckg} shows the projected sensitivities to the DM annihilation rate
assuming that the background at high energies is given by the predictions in Ref.~\cite{Zhou:2016ljf},\footnote{%
There are other sources of background, as for instance a $\gamma$-ray astrophysical diffuse emission and residuals from cosmic-ray events, which we do not consider here.} which is below detector sensitivity. 
Since at present, the HERD Collaboration does not provide a differential sensitivity for continuum emission,
the corresponding curve for HERD is estimated on the basis of the sensitivity for line searches in the Galactic center reported 
in Ref.~\cite{Huang:2015fca} approximating the annihilation spectra by its sharp peak 
(in the $E^2\rmd N/\rmd E$ representation) at the end point (neglecting contributions from lower energies). 
The resulting sensitivity exceeds the one of the Fermi-LAT full mission at masses above $\sim 400\,$GeV, in rough agreement with the expectation mentioned above.
Note that it is likely that if HERD performed solar observations, the sensitivity could be better as we expect the background levels to be smaller compared to the Galactic center.

We find that some of the benchmark points ({\it i.e.}~2a and 3a) in the $m_X \gtrsim 100\, \rm GeV$ range of our simplified model will likely be probed by HERD with roughly 5 years exposure. Similar to HERD, LHAASO will be able to probe our simplified model benchmarks with $m_X~\gtrsim~1~\TeV$ and hence provide information complementary to the DM searches with CTA and XENON1T. Compared to the conservative estimate of the HERD sensitivity, LHAASO displays superior sensitivity in the range of $m_X \gtrsim 500 \GeV$. 

While all benchmark points above $m_X \gtrsim 100$ GeV are within the reach of XENON1T or LZ (\emph{cf.}~Fig.~\ref{fig:dd}) they fall into two classes regarding indirect detection prospects:

\emph{(i)}~Benchmark points 4b and 5b are well outside the reach of CTA (\emph{cf.}~Fig.~\ref{fig:dwarflim}). Hence,
solar $\gamma$-rays are a unique probe of the self-annihilating nature of DM in this parameter region\@.
Note also that for these points capture and annihilation proceeds out of equilibrium (see Tab.~\ref{tab:fluxes}).
In this case, together with an independent measurement of the DM-nucleon scattering cross section from
direct detection experiments information about the annihilation cross section could be gained.

\emph{(ii)}~Benchmark points 2a--5a are on the edge of the reach of CTA (\emph{cf.}~Fig.~\ref{fig:dwarflim}).
However, solar $\gamma$-ray measurements could offer useful complementary information to a possible Galactic center observation. Namely, an observation of a $\gamma$-ray signal in the Galactic center will always be plagued by uncertainties in the DM density profile, as well as possible unaccounted-for background sources. An additional observation of a $\gamma$-ray signal coming from the Sun would provide a smoking gun confirmation of such a signal.
It would also indicate that DM annihilation proceeds via a long-lived mediator -- a piece of information which would be impossible to infer from the observations of the Galactic center.

For  DM below 100 GeV solar $\gamma$-rays observations are most likely not  sensitive to our model, as the expected flux is below the background (by almost two orders of magnitude for point 1a), unless  angular resolution can significantly improve the signal to noise ratio. This region of parameter space will, however, be probed by (conventional) direct or indirect detection experiments.
Benchmark point 1a is very close to the current sensitivity of Fermi-LAT targeting dwarf spheroidal galaxies while its DM-nucleon scattering cross section is below the neutrino background.  In contrast benchmark points 1b, 2b and 3b will be efficiently probed by direct detection experiments while
$\gamma$-ray observations of the Galactic center or dwarf spheroidal galaxies do not provide useful constraints.

Finally, we note that there exists a part of the parameter space which will remain unexplored by future (in)direct detection probes but will likely be probed by the next generation of beam dump experiments, such as NA62~\cite{Hahn:1404985} and ShiP~\cite{Alekhin:2015byh}.

\section{Conclusions}\label{sec:concl}

The Sun represents a potential nearby reservoir of DM accumulated in its center. Since it is a poor source of high-energy $\gamma$ rays, the Sun becomes an interesting target for DM studies using the next generation of cosmic-ray and $\gamma$-ray detectors.

In this paper we studied the possibility of probing DM particle properties via the observation of $\gamma$-ray signals from the Sun. Such signals arise from DM annihilation into a pair of long-lived mediators that consequently decay outside the solar surface, before reaching the Earth. As an illustration we considered a simplified model extending the SM by a Dirac DM particle and a mixed scalar-pseudoscalar mediator.  The above signature becomes relevant for small mediator masses and small couplings between the mediator and the SM quarks. This part of the parameter space is not challenged by the LHC and can easily accommodate the relic density via thermal freeze-out while evading current direct detection bounds. We showed that the parameter space cornered by beam dump experiments and BBN constraints (towards large and small mediator couplings, respectively) exhibits a large overlap with sufficiently long mediator lifetimes to decay between the solar surface and the Earth's orbit.
Considering model points that are within the reach of the next generation of direct detection experiments, we demonstrated that the prospects for $\gamma$-ray observations of the Sun can provide compatible or superior sensitivity compared to the observations of the Galactic center or dwarf spheroidal galaxies. In particular we found that the DM induced solar $\gamma$-ray  flux can be 
up to five orders of magnitude larger than the total flux expected from dwarf spheroidal galaxies
which are currently among the most sensitive probes of DM\@.
Moreover, for heavy DM (\emph{e.g.}~$m_X \gtrsim 500\,\text{GeV}$) the Sun is essentially a background free environment for $\gamma$-ray observations, making it a very attractive target for TeV scale DM searches. 

Our results, although presented in the context of a particular simplified model, are easily extended to other theoretically motivated DM scenarios and represent a proof of principle that solar $\gamma$-ray observations provide a unique or complementary probe of $\gtrsim 1 \TeV$ scale DM with long-lived mediators. If  DM is a Dirac fermion carrying a Peccei-Quinn charge and is connected to the SM via at least a complex scalar field, there will be a region of the model parameter space featuring a long-lived mediator and hence solar $\gamma$-ray signatures. Such ``axion-mediated'' models have previously been considered in Refs.~\cite{Batell:2009zp,Lee:2012bq,Ibarra:2013eda}. 

Remarkably, even though the process of DM capture and annihilation does not readily fulfill the equilibrium conditions in our simplified model, it is possible to produce solar $\gamma$-ray fluxes of order $(10^{-12} \!-\!10^{-9}) \, {\rm cm}^{-2} \rm s^{-1}$, and within the reach of future HERD and LHAASO experiments. Notice that the equilibrium condition is typically violated for viable thermal relic DM models in which the main capture process is via spin-independent elastic scattering, due to strong LUX bounds.
The lack of equilibrium also implies that within the framework of our simplified model, measurement of the solar $\gamma$-ray signal would provide useful information on the DM annihilation cross section when used in conjunction with an observed direct detection signal. Observation of a solar $\gamma$-ray signal would also indicate the existence of a long-lived mediator, information which could not be inferred from a detection in the Galactic center and/or in direct detection experiments.

\begin{acknowledgments}

The authors acknowledge A. Martini for his help in the early stage of this paper and the anonymous referees for a revision that has improved the manuscript. We also thank F. Calore, A. Cuoco, M. Cirelli, C.~Delaunay, D.~Guadagnoli,
X. Huang and B.~L\"ulf for useful discussions and comments, as well  as N. Ozak Munoz for useful discussions about solar $\gamma$-ray backgrounds. C.A. acknowledges the support of the ATTRACT 2015 - Brains for Brussels Innoviris grant. M.B. is funded by the MOVE-IN Louvain, Marie Curie cofund grant. J.H. acknowledges support by the German Research Foundation DFG through the research unit ``New physics at the LHC''. M.L. acknowledges support by the Fonds de la Recherche Scientifique - FNRS under Grant No. IISN 4.4512.10.
\end{acknowledgments}

\bibliography{bibliofile}

\end{document}